**ONSET OF THE MAGNETIC EXPLOSION IN SOLAR POLAR CORONAL X-RAY JETS**

Short Title: **MAGNETIC EXPLOSION IN SOLAR CORONAL JETS**

Ronald L. Moore[1,2], Alphonse C. Sterling[1], Navdeep K. Panesar[1]

[1]Heliophysics and Planetary Science Office, ST13, Marshall Space Flight Center, Huntsville, AL 35812, USA

ron.moore@nasa.gov

[2]Center for Space Plasma and Aeronomic Research (CSPAR), University of Alabama in Huntsville, Huntsville, AL 35805, USA

**ABSTRACT**

We follow up on the Sterling et al (2015) discovery that nearly all polar coronal X-ray jets are made by an explosive eruption of closed magnetic field carrying a miniature filament in its core. In the same X-ray and EUV movies used by Sterling et al (2015), we examine the onset and growth of the driving magnetic explosion in 15 of the 20 jets that they studied. We find evidence that: (1) in a large majority of polar X-ray jets, the runaway internal/tether-cutting reconnection under the erupting minifilament flux rope starts after both the minifilament's rise and the spire-producing external/breakout reconnection have started; and (2) in a large minority, (a) before the eruption starts there is a current sheet between the explosive closed field and the ambient open field, and (b) the eruption starts with breakout reconnection at that current sheet. The variety of event sequences in the eruptions supports the idea that the magnetic explosions that make polar X-ray jets work the same way as the much larger magnetic explosions that make a flare and coronal mass ejection (CME). That idea, and recent observations indicating that magnetic flux cancelation is the fundamental process that builds the field in and around the pre-jet minifilament and triggers that field's jet-driving explosion, together suggest that flux cancelation inside the magnetic arcade that explodes in a flare/CME eruption is usually the fundamental process that builds the explosive field in the core of the arcade and triggers that field's explosion.

# 1. INTRODUCTION

Sterling et al (2015) recently found that practically all coronal X-ray jets that happen in the Sun's polar coronal holes are produced by the eruption of a small filament (which they called a minifilament) of cool (T = $10^4$ – $10^5$ K) plasma from the base of the jet. In the onset of each of 20 random jets observed in movies of polar coronal holes taken in coronal X-ray emission by the X-Ray Telescope (XRT) on *Hinode*, from a simultaneous movie taken in coronal extreme-ultraviolet (EUV) emission by the Atmospheric Imaging Assembly (AIA) on the *Solar Dynamics Observatory (SDO)*, they found an erupting minifilament in the jet's base. From the observed involvement of the minifilament eruptions in the production of the jets, they surmised that the erupting minifilament is carried in the erupting sheared core field of an initially-closed small exploding bipolar magnetic arcade seated at an edge of the jet's base. In their interpretation, the minifilament eruption is a manifestation of a magnetic explosion that is a miniature version of the big magnetic explosions that drive a flare with or without a coronal mass ejection (CME) and typically have an erupting filament inside the exploding field, big solar eruptions like the ones presented in Moore (1988), Moore et al (2001) and Moore & Sterling (2006).

The three schematic drawings in Figure 1 are from Sterling et al (2015). They depict the progression of the process – as inferred by Sterling et al (2015) from their observations – by which a minifilament eruption produces a jet. These drawings are for the case of a jet in a coronal hole in which the polarity of the majority of the flux, the polarity of the open field that fills the coronal hole, is negative. Each drawing is a sketch of a few representative field lines in and near the minifilament, field lines viewed in projection on a vertical cross-sectional plane through the middle of the minifilament and orthogonal to the polarity inversion line below the pre-eruption minifilament, so that the minifilament is viewed end-on. The cross-sectional plane also cuts through the middle of a clump of minority-polarity (positive) flux sitting immediately left of the pre-eruption minifilament. Along the rightward edge of this island of positive-polarity flux, the polarity inversion line that is the periphery of the island lies below and traces the minifilament. All of the flux in the island is in the positive-polarity feet of closed field that reaches over the polarity inversion line and connects to nearby negative flux. The closed field fills the base of the jet.

The first (left) drawing in Figure 1 shows the gradually-evolving field configuration a minute or so before the minifilament starts to erupt. By this time, the field's prior evolution, driven by magnetoconvection flows in and below the photosphere, has resulted in the lobe of closed field on the right side of the positive-polarity island being more compact than the lobe of closed field on the left side, and more contorted from its potential-field configuration. In this drawing, only the greatly sheared and twisted field in the core of the right-side lobe is sketched; the relatively less contorted outer envelope of this lobe is not drawn. The core field of the more-compact lobe is sheared and twisted enough for it to hold a minifilament in it, and for it to have enough free magnetic energy that the lobe is capable of exploding.

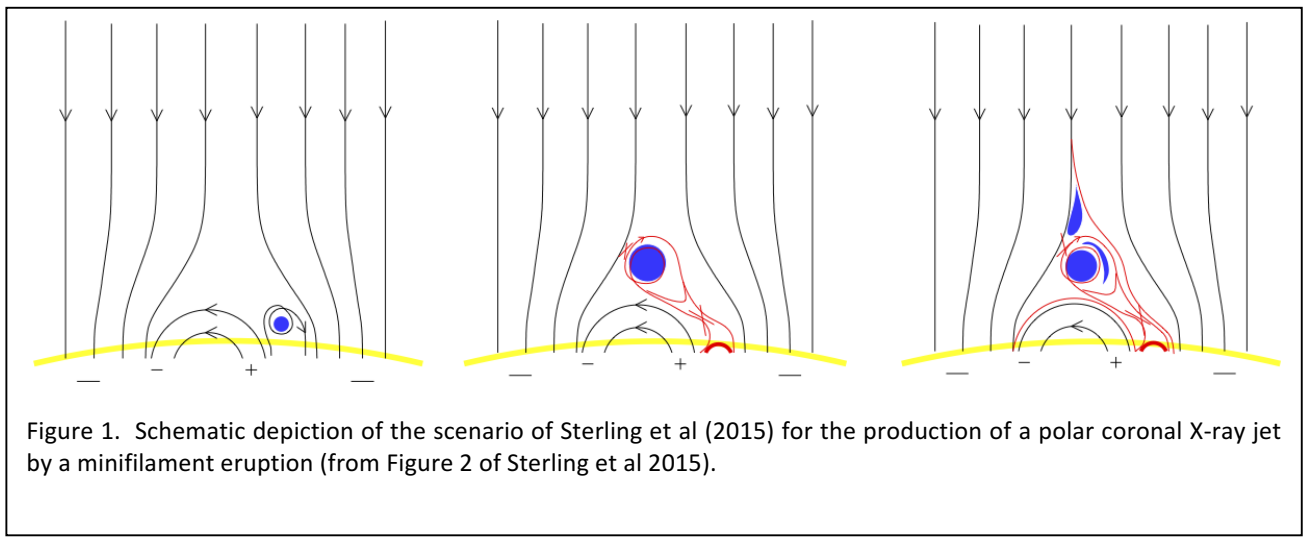

Figure 1. Schematic depiction of the scenario of Sterling et al (2015) for the production of a polar coronal X-ray jet by a minifilament eruption (from Figure 2 of Sterling et al 2015).

In the first drawing in Figure 1, the field in the larger, more-nearly-potential lobe is opposite in direction to the open field that is bent toward it and that is rooted to the right of the lobe having the minifilament in it, and field in the envelope of the lobe having the minifilament is opposite in direction to the open field that is bent toward it and that is rooted to the left of the more-nearly-potential lobe.  So, in the field configuration sketched in the first drawing, somewhere in the region above the closed field and below the canopy of open field over the closed field, in the blank region where no field lines are drawn, there is a magnetic null point or current sheet that is not shown in this drawing.

The second (middle) drawing in Figure 1 depicts the minifilament eruption well underway a few minutes after the start of the eruption.  In the eruption scenario envisioned in Sterling et al (2015), a minute or so into the time interval between the two times depicted in the first and second drawings of Figure 1, the minifilament-holding lobe is rendered unstable – by some unspecified mechanism – and starts erupting upward along the outside of the envelope of the adjacent lobe of closed field in the base of the forthcoming jet.  A tacit assumption of the scenario of Sterling et al (2015) is that as the minifilament-carrying lobe erupts its envelope field compresses the field in the magnetic-null region between it and the open field on the opposite side, deforming the null-region field to make and/or enlarge a current sheet between the envelope of the erupting lobe and the open field that the erupting lobe is approaching.  The jet spire results from reconnection driven at that current sheet by the erupting lobe, reconnection of the outside of the erupting closed field with the impacted open field.  This external reconnection of the erupting lobe simultaneously produces both new open field lines (along which the plasma making up the jet spire is ejected) and new closed field loops (filled with reconnection-heated plasma) that are added to the outside of the non-erupting lobe of closed field. The second drawing in Figure 1 depicts the eruption at the very start of the spire-producing external reconnection of the envelope of the erupting lobe, when the external reconnection has as yet produced no appreciable new open field and no appreciable new closed field.

In the scenario of Sterling et al (2015), besides external reconnection of the erupting lobe there is also internal reconnection of the two opposite-polarity legs of the erupting lobe. In the second drawing in Figure 1, the red field lines depict field that has been reconnected by the internal reconnection. The reconnected field released upward by the internal reconnection wraps around the flux rope that threads and carries the minifilament. The reconnected field released downward by the internal reconnection makes a miniature flare arcade, depicted here by the low thick red loop over the polarity inversion line from which the minifilament erupted. In this scenario, the miniature flare arcade is the bright point (called the "jet-base-edge bright point" or the "jet bright point" (JBP)) typically seen at the edge of the base of coronal X-ray jets in coronal X-ray images (e.g., Savcheva et al 2009, Sterling et al 2015). The hot-plasma loops added to the outside of the non-erupting lobe of the jet base by the external reconnection of the erupting lobe are interpreted as the brightening seen in the interior of the jet base in coronal X-ray images, brightening that is offset from and distinct from the JBP at the edge of the jet base (as in the example jets presented in the present paper).

The Sterling et al (2015) scenario is independent of and does not specify the answers to the following two questions about the events in the onset and growth of the explosive lobe's eruption: (1) what triggers the eruption, and (2) whether the external reconnection starts before, simultaneously with, or after either the internal reconnection or the rise of the minifilament start. Sterling et al (2015) mention that a likely trigger of the eruption is magnetic flux cancelation in the feet of the magnetic field in and around the minifilament flux rope, and Panesar et al (2016, 2018) have presented compelling evidence that in most coronal jets in quiet regions and coronal holes flux cancelation at the polarity inversion line under the pre-jet minifilament triggers the eruption. The results of Panesar et al (2016, 2018) are from 23 randomly selected jets, and are consistent with many recent studies of smaller samples of jets in which there is evidence for flux cancelation being the trigger (e.g., Young & Muglach 2014a, 2014b; Chen & Innes 2016; Chen et al 2017; Zhu et al 2017; Chandra et al 2017). The present paper is a direct follow-on to the Sterling et al (2015) paper. We investigate the above question (2) by closely discerning the first signs of the external reconnection, the internal reconnection, and the rise of the minifilament along the outside of the stable lobe, in 15 of the 20 polar X-ray jets studied in Sterling et al (2015). This yields new evidence for the magnetic explosions that drive coronal jets being miniature versions of the magnetic explosions that drive major flares and CMEs.

Following Moore & Roumeliotis (1992), for the driving magnetic-explosion eruptions in flare/CME events and in coronal jets, we consider the trigger of the eruption to be distinct from the onset or start of the eruption. We take the trigger of the eruption to be the last bit of the explosive magnetic lobe's pre-eruption evolution that renders the lobe (the sheared-core magnetic arcade) eruptively unstable. We take the onset or start of the eruption to be part of the eruptive process itself, part of the process by which the eruption grows and becomes progressively faster. In this view, in coronal jets, once the photospherically-driven gradual evolution of the minifilament-carrying magnetic lobe renders the lobe unstable to eruption, then, depending of the specifics of the field configuration, either the minifilament eruption starts on its own via ideal MHD instability (or loss of magnetostatic equilibrium) before the start of any eruption-driven reconnection (internal or external), or the driven internal reconnection or the driven external reconnection or both start simultaneously with the minifilament eruption. In summary, even when either reconnection depicted in Figure 1 starts simultaneously with the rise of the minifilament, we consider the start of that reconnection to be an integral part of the eruption, not the trigger of the eruption.

The third (right) drawing in Figure 1 depicts the eruption a few minutes after the time depicted in the second drawing. By the time depicted in the third drawing, the external reconnection has eaten through the erupting lobe's envelope, opening it to produce new open field (represented in the third drawing by the new red open field line) and new closed field (represented by the new red outer loop added to the non-erupting lobe), and is now opening the minifilament-carrying core of the erupting closed field, resulting in cool minifilament plasma being driven out along the newly-open field threading that plasma. Also in the eruption scenario depicted in Figure 1, from before the time of the second drawing and continuing beyond the time of the third drawing, the internal reconnection of the legs of the erupting field continuously builds the JBP.

In coronal X-ray movies from *Hinode*/XRT, a great majority (> 90%) of coronal X-ray jets in polar coronal holes are seen to be of one or the other of two morphologically different classes, called standard jets and blowout jets; only a small minority (< 10%) have ambiguous morphology that is neither decidedly that of standard jets nor decidedly that of blowout jets but a mixture of both (Moore et al 2010, 2013). In coronal X-ray movies, the most striking morphological difference between standard jets and blowout jets is that in standard jets the width of the spire throughout its life is much less than the width of the jet base, whereas in blowout jets the spire grows in width to become about as wide as the base. Movies of polar coronal X-ray jets from *SDO*/AIA in AIA's He II 304 Å band show that (1) in a large majority (~ 80%) of standard jets the spire has no noticeable cool (T ~ $10^5$ K) plasma component, whereas in large majority (~ 90%) of blowout jets the spire has an obvious cool component, and (2) in standard-jet spires that do have a cool component the spire's cool component shows little if any lateral expansion, whereas in blowout jets the spire's cool component shows obvious lateral expansion, obvious growth in width (Moore et al 2013).

Based on coronal X-ray and EUV movies of polar coronal X-ray jets, the minifilament-eruption scenario of Sterling et al (2015) for jet production offers the following interpretation of the standard-jet/blowout-jet dichotomy. Many standard jets are plausibly produced when the eruption of the minifilament lobe is mostly a confined eruption, an eruption similar to the larger magnetic-field eruptions that produce a filament eruption and flare but no CME. That is, in a standard-jet eruption, the minifilament-carrying core of the erupting lobe is plausibly largely confined and arrested within the close-field base of the jet so that the external reconnection stops before any or more than a little of the field threading the erupting minifilament is opened. That would plausibly result in a narrow jet spire that does not grow much in width and usually has no cool component. A blowout jet is produced when the eruption of the minifilament-carrying lobe is a blowout eruption as in filament eruptions in which the filament-carrying erupting field blows out to produce a CME in tandem with a flare. In a blowout-jet eruption, instead of becoming a miniature magnetic-bubble CME that escapes into the solar wind, as the erupting lobe explodes out of the jet base, most of the lobe – including most of or all of the minifilament-carrying erupting core field – is opened by external reconnection that becomes much more widespread than in standard jets. That results in an X-ray-emitting hot spire that grows in width as the minifilament becomes a wide cool-plasma spray made by the external reconnection opening the field threading the minifilament.

The external reconnection in the minifilament-eruption scenario for the production of polar coronal X-ray jets corresponds to the "breakout" reconnection of Antiochos (1998) and Antiochos et al (1999). Breakout reconnection occurs in certain of the magnetic eruptions that make a flare and CME. Breakout reconnection is the external reconnection that occurs during the eruption of a sheared-core magnetic arcade that – prior to eruption – is seated in a multipolar magnetic field in which overlying the explosive arcade there is a cap field that is opposite in direction to the arcade field. So, "breakout" external reconnection of the envelope of the arcade can occur at the interface of the envelope with the cap. The eruption of the explosive arcade can begin with the onset of such runaway reconnection at the interface. If the arcade's eruption begins some other way, soon after the eruption begins, breakout reconnection is driven at the interface by the erupting arcade and contributes to the unleashing and growth of the explosion (Moore & Sterling 2006).

Moore & Sterling (2006) review observations and MHD simulations of the onsets of filament-carrying sheared-core magnetic-arcade eruptions that make a flare and CME. For eruptions in which breakout reconnection of the erupting arcade can occur, they describe three different mechanisms that can initiate the eruption. Depending on the specifics of the pre-eruption structure and evolution of the magnetic field in and around the arcade, the eruption can be initiated by any one of these mechanisms acting alone, or by any two acting together, or by all three acting together.

Moore & Sterling (2006) call one eruption-initiation mechanism "runaway internal tether-cutting reconnection." In eruptions started by this mechanism alone, the eruption of the filament-carrying sheared-field core of the arcade begins simultaneously with the start of runaway reconnection of impacted opposite-polarity legs of sheared field that closely envelops the flux rope in which the filament is suspended. The legs of the enveloping sheared field tether the filament flux rope down to the photosphere, balancing the upward push of the magnetic pressure. When the reconnection starts, the filament flux rope is immediately less tied-down and starts rising up. More opposite-polarity field legs then collapse together below the rising filament flux rope, speeding up the tether-cutting reconnection, which in turn results in the flux rope rising more, and so forth. The flux-rope rising and the reconnection are thus coupled in a positive-feedback runaway process that explosively grows the eruption (Moore & LaBonte 1980; Moore & Roumeliotis 1992; Moore et al 2001). Many sheared-core-arcade eruptions that have no access to breakout reconnection have been observed to start in this manner (e.g., Moore et al 2001). In a recently-studied sheared-core-arcade eruption that underwent breakout reconnection as it erupted to produce a flare and CME, emission from plasma heated by runaway internal tether-cutting reconnection of the legs of the erupting sheared core field started several minutes before the first discernible emission from plasma heated by the breakout reconnection (Joshi et al 2017). Apparently, this eruption was initiated by runaway internal tether-cutting reconnection and soon after drove breakout reconnection of the outer envelope of the erupting arcade.

Another of the three mechanisms described by Moore & Sterling (2006) for initiating the eruption of a sheared-core magnetic arcade is "runaway external tether-cutting reconnection," the breakout reconnection mentioned above. In an eruption that starts by this mechanism alone, there is a current sheet on the outside of the pre-eruption arcade, and the eruption of the arcade starts with the onset of runaway reconnection at this external current sheet. In concert with the legs of the sheared field in the core of the arcade, the outer loops of the arcade help tether the sheared core field to the photosphere: some of the upward push of the magnetic pressure of the sheared field in the interior of the arcade is balanced by the downward pull of the field in the legs of the arcade's outer loops. The external reconnection of the outer loops cuts some of the arcade's tethers to the photosphere. When the reconnection at the external current sheet starts, the sheared field in the interior of the arcade is immediately less tied-down. When the signal of the tether cutting by the onset of the external reconnection, traveling at the Alfven speed, reaches the core, the filament flux rope in the core starts rising up. The less-tethered interior field then pushes the arcade's outer loops harder against the external current sheet. That drives more breakout reconnection, which further unleashes the filament-carrying erupting sheared-core arcade, and so forth, in a positive-feedback runaway process that explosively grows the eruption (Antiochos 1998; Antiochos et al 1999; Moore & Sterling 2006). Eruptions that appear to be initiated this way have been observed, with EUV and/or coronal X-ray emission from plasma heated by breakout reconnection starting at the onset of the eruption of the sheared-field core of the arcade (e.g., Aulanier et al 2000; Sterling & Moore 2004a; Li et al 2005).

The third eruption-initiation mechanism described by Moore & Sterling (2006) is "ideal MHD instability or loss of equilibrium." In an eruption initiated by this mechanism alone, there is no runaway reconnection, internal or external, initially occurring in the onset of the eruption. Instead, the filament-carrying flux rope starts rising on its own, as a result of having been rendered MHD unstable or out of magnetostatic equilibrium by photospherically-driven gradual evolution of the field in and around the arcade. Soon after the onset of an eruption that begins this way but has access to external reconnection, breakout reconnection and runaway internal reconnection set in and become part of the explosive eruption process. The filament eruption in the sheared-core-arcade eruptions that make a CME in tandem with a flare typically begins with a slow-rise phase in which the rise speed has little or no acceleration; eventually the eruption accelerates and transitions to its fast-rise phase shortly before or during the flare's impulsive phase (e.g., Sterling & Moore 2005; Moore & Sterling 2006). Whether or not the erupting arcade has access to external reconnection, the eruption's slow-rise phase often starts well before there is any discernible emission from heating by either breakout reconnection or runaway internal tether-cutting reconnection (e.g., Sterling & Moore 2003; Sterling & Moore 2004b). In these cases, initiation of the eruption by ideal MHD instability or loss of equilibrium possibly starts the eruption by acting alone, but usually it cannot be ruled out that from the very start of the eruption the initiation partly or entirely results from breakout reconnection and/or runaway internal tether-cutting reconnection that is too weak to produce discernible heating (Sterling & Moore 2003; Moore & Sterling 2006).

For eruptions that produce a flare and CME, observations of the eruption onsets presented in the papers cited above show that, among sheared-core-arcade eruptions that have access to external reconnection, the start of the filament's rise can either lead or coincide with the start of breakout emission. Because the minifilament eruptions in their 20 jets have the same character as the larger-scale filament eruptions that produce a flare and CME, including starting out slowly rising before transitioning to a much faster outward explosion, Sterling et al (2015) infer that the minifilament eruptions in polar coronal X-ray jets are miniature versions of the filament eruptions that produce a flare and CME. If that inference is true, it is reasonable to expect that the start of the minifilament's rise should have the same diversity of timing relative to the start of the eruption's breakout emission as the filament eruptions in sheared-core-arcade eruptions that produce a flare and CME. That is, we should expect that (1) the minifilament starts erupting before any sign of runaway reconnection (internal or external) in some polar X-ray jet eruptions, but not all, (2) in some the runaway internal tether-cutting reconnection starts simultaneously with the start of the minifilament eruption, and (3) in some the breakout reconnection starts simultaneously with the start of the minifilament eruption.

The present paper presents new evidence that the minifilament eruptions in polar coronal X-ray jets are miniature versions of the filament eruptions in the magnetic explosions that produce a flare and CME. We examine 15 of the 20 jets studied in Sterling et al (2015), the 15 jets in which the onset of the JBP was observed in the XRT movie (the onset of the JPB was not eclipsed by the limb or missed in a time gap in the movie). For each jet we have identified the time of the first detection of the rising minifilament in the jet's AIA movie made by Sterling et al (2015). In the XRT movie of each jet, we have identified the time of the first detection of emission from breakout reconnection and the time the first detection of emission from internal tether-cutting reconnection (the start time of the JBP). The observed leads and lags between pairs of these three times mimic corresponding leads and lags observed in onsets of the magnetic explosions that make a flare and CME. This similarity supports the view that the magnetic explosions that make a polar X-ray jet are miniatures of the magnetic explosions that make a flare and CME.

## 2. DATA

### 2.1. Fifteen Random Polar Coronal X-Ray Jets

The 15 polar coronal X-ray jets studied in this paper are numbered and listed chronologically in Table 1. For each jet, we have the *Hinode*/XRT movie in which the jet was originally found for the Moore et al (2013) study of polar coronal X-ray jets that were observed both in an XRT movie of a polar coronal hole and in the full-disk EUV movies from *SDO*/AIA. In that study, from the jet's form in the XRT movie, we judged each jet to be a standard jet or a blowout jet, or an ambiguous mix of those two morphological types. In our **15** jets, only one (Jet 9) has ambiguous morphology, three jets (Jets 3, 4, and 5) are standard jets, and the other 11 jets are blowout jets. The XRT movie of each jet was taken with the Ti-Poly filter. Except for Jet 15, the cadence of the XRT movie is 30s; for Jet 15, the cadence is 60 s. To better detect the onsets of coronal X-ray brightenings in each jet, we have made from the XRT movie the corresponding base-difference movie for which the first frame of the XRT movie is the base image. Subtracting the base image from the image in each frame of the XRT movie yields the base-difference movie. In each frame of the base-difference movie, medium-gray pixels are those that in the corresponding frame of the XRT movie have the same brightness as in the first frame of the XRT movie, lighter-gray and white pixels are those that are brighter than in the first frame, and darker-gray and black pixels are those that are dimmer than in the first frame.

Table 1. Our Polar X-ray Jets with Times of Events Observed in Them by *Hinode*/XRT or *Solar Dynamics Observatory*/AIA

| Jet Number[*] | Date (2010) | Jet Type | AIA Channel (Å) | First Det. MF Rise AIA (UT) | First Det. JBP XRT (UT) | First Det. BIB XRT (UT) | First Det. Spire XRT (UT) | First Det. Spire AIA (UT) | Spire Max XRT (UT) | Spire Max AIA (UT) |
|---|---|---|---|---|---|---|---|---|---|---|
| 1 (1) | Jul 24 | Blowout | 211 | 15:51:48 | 15:56:08 | 15:51:38 | 15:58:08 | 15:58:12 | 16:07:08 | 16:03:00 |
| 2 (3) | Aug 26 | Blowout | 193 | 14:03:54 | 14:09:51 | 14:08:21 | 14:10:51 | 14:10:42 | >14:15:51[1] | 14:18:42 |
| 3 (4) | Aug 27 | Standard | 193 | 11:27:06 | 11:34:28 | 11:31:58 | 11:33:28 | 11:44:42 | 12:03:01 | 12:02:42 |
| 4 (5) | Aug 27 | Standard | 193 | 11:30:00 | 11:38:58 | 11:34:28 | 11:42:58 | 11:43:54 | 11:50:31 | 11:49:30 |
| 5 (6) | Aug 28 | Standard | 211 | 11:33:00 | 11:40:46 | 11:38:46 | 11:40:46 | 11:42:12 | 11:46:16 | 11:46:12 |
| 6 (8) | Sep 5 | Blowout | 211 | 21:09:12 | 21:17:18 | 21:13:18 | 21:18:48 | 21:17:36 | 21:26:50 | 21:21:36 |
| 7 (9) | Sep 8 | Blowout | 193 | 01:22:18[2] | 01:28:06 | 01:21:06 | 01:28:36 | 01:27:06 | 01:35:36 | 01:35:30 |
| 8 (10) | Sep 9 | Blowout | 304 | 20:03:66 | 20:16:42 | 20:07:12 | 20:12:42 | 20:19:08 | 20:18:12 | 20:22:44 |
| 9 (11) | Sep 9 | Ambiguous | 211 | 20:16:54[3] | 20:24:43 | 20:11:12 | 20:20:42 | 20:20:36 | 20:28:45 | 20:27:36 |
| 10 (12) | Sep 9 | Blowout | 193 | 21:56:30 | 22:03:50 | 22:01:20 | 22:04:50 | 22:04:06 | 22:11:20 | 22:11:18 |
| 11 (14) | Sep 9/10 | Blowout | 304 | 23:58:56 | 23:59:28 | 00:00:28 | 23:59:28 | 00:02:08 | 00:04:08 | 00:06:08 |
| 12 (15) | Sep 11 | Blowout | 193 | 00:35:30 | 00:37:30 | 00:36:00 | 00:38:00 | 00:35:54 | 00:42:05 | 00:37:30 |
| 13 (17) | Sep 17 | Blowout | 211 | 20:30:06[4] | 20:33:42 | <20:11:10[5] | 20:23:10 | 20:41:00 | 20:49:12 | 20:50:36 |
| 14 (18) | Sep 17 | Blowout | 193 | 22:04:06 | 22:07:18 | 22:04:47 | 22:09:48 | 22:07:18 | 22:12:18 | 22:12:54 |
| 15 (19) | Sep 27 | Blowout | 193 | 00:32:18 | 00:38:28 | 00:37:28 | 00:36:28 | 00:37:30 | 00:40:33 | 00:42:42 |

[*] The number in parentheses is the number of the same jet in Sterling et al (2015).

[1] This is the time of the start of a time gap of 7.5 minutes in the XRT movie; spire maximum occurred during the gap.

[2] Faint nebulous precursory upflow along minifilament-side of base arch became discernible at 01:17:54 UT.

[3] Faint nebulous precursory upflow along minifilament-side of base arch became discernible at 20:14:12 UT.

[4] Faint nebulous precursory upflow along minifilament-side of base arch became discernible at about 20:00 UT.

[5] This is the time of the XRT movie's start, which caught the base interior brightening in progress.

We also have for each of our 15 jets an AIA EUV movie that was used in the Sterling et al (2015) study of erupting minifilaments in polar coronal X-ray jets.  For each jet, the AIA movie that we use is from the AIA channel in which, in the Sterling et al (2015) study, the erupting minifilament was found to have the best visibility.  The channel of the AIA movie for each jet is given in Table 1.  Depending on the jet's surroundings and proximity to the limb in the coronal hole, the erupting minifilament is most visible in one of three channels: the 193 Å channel, the 211 Å channel, or the 304 Å channel.  In the jets in which the erupting minifilament is most visible in the 304 Å channel, the jet's base is not right at the limb; instead, the base, the erupting minifilament, and the jet spire are all viewed against the disk inside the limb.  In polar coronal X-ray jets, the 193 Å, 211 Å, and 304 Å channels detect emission from plasmas at temperatures around $1.5 \times 10^6$ K, $2 \times 10^6$ K, and $5 \times 10^4$ K, respectively (Lemen et al 2012).  At the onset of its eruption the minifilament is seen as a dark absorption feature in the images from any of these channels, indicating that the temperature of the minifilament plasma is in the temperature range of the chromosphere and cooler transition region ($5 \times 10^3$ K < T < $5 \times 10^4$ K).  The cadence of the images from the AIA channels that we use is 12 s.  To enhance the visibility of the minifilament, each frame of the AIA movies used in Sterling et al (2015) and used again here is the sum of two consecutive frames of the full-cadence AIA movie.  Consequently, the cadence of the AIA movies that we use is 24 s instead of 12 s.

For each jet, the XRT movie shows in coronal X-ray emission the onset and evolution of the spire and the brightening of the base.  The brightening of the base has two spatially-separate components: the JBP at an edge of the base and brightening in the interior of the base, which we call the "base interior brightening" (BIB).  We take the spire and the BIB to be signatures of the external reconnection of the minifilament-carrying erupting magnetic arcade that drives the production of the jet.  We take the JBP to be a signature of internal reconnection of opposite-polarity legs of the erupting arcade.  By stepping through the XRT base-difference movie of each jet, we have identified the start times of the JBP, the BIB, and the spire in the XRT movie.  We have taken the start time (the time of first detection) of each of these features to be the time of the frame of the base-difference movie in which the brightening of the feature first becomes visually discernible.  For each jet, the identified start times of these features in the XRT movie are given in Table 1.  The identification of the start-time frame is usually uncertain by ± 1 frame around the selected frame.  So, the uncertainty in the start time of each feature in the XRT movie is roughly ± 30 s for Jets 1-14 and roughly ± 60 s for Jet 15.  For each jet, the time of spire maximum in the XRT movie is given in the next-to-last column of Table 1; that time is the time of the visually selected frame of the XRT movie in which we judge the spire to be most prominent in terms of brightness and spatial extent combined.  In the same way as for the feature start times in the XRT movie, the uncertainty in the selected time of spire maximum in the XRT movie is roughly ± 30 s for Jets 1-14 and roughly ± 60 s for Jet 15.

For each jet, by stepping through the AIA movie, we have visually identified the first frame in which the minifilament is unambiguously seen to have started rising. The time of this frame is the time we take for the start (the first detection) of the minifilament (MF) rise in the AIA movie, the time given in the column in Table 1 labeled: First Det. MF Rise AIA. By stepping through the AIA movie of each jet, we also visually identified the start time of the spire and the time of spire maximum in the AIA movie, the time of the frame in which the spire is most prominent in terms of brightness and spatial extent combined. These two times are also listed in Table 1. In the same way as for the uncertainty in the four times identified from the XRT movie, the uncertainty in each of the three times identified from the AIA movie is roughly ± 24 s for each jet. We do not identify the start of the JBP in the AIA movie because, during the birth of the JBP in the XRT movie, the JBP is often not yet visible in the AIA movie because it is hidden behind foreground chromospheric and cool-transition-region plasma that blocks the coronal EUV emission from the young JBP.

Our main finding is the variation among our 15 jets of the first signature of the eruption onset. We find this by determining for each jet, from the times given for that jet in Table 1, the eruption onset's first signature: What starts (is detected) first, the minifilament's rising, the BIB and spire, or the JBP, or do two of these or all three start simultaneously to within the uncertainty of the timing?

*2.2. Example Jet Eruptions*

*2.2.1. Jet 5*

Our first example jet eruption is that of Jet 5 of Table 1. Jet 5 is a typical standard jet in that in the X-ray images its spire is much narrower than the base of the jet, is single-stranded (does not have two or more separate strands as blowout jets typically have), and is fainter than the spire in typical blowout jets. The eruption of Jet 5 is an example of a jet eruption in which the minifilament's rising is detected in the AIA movie before any coronal X-ray emission from plasma heated by either internal or external reconnection is discernible in the XRT movie.

The erupting minifilament in Jet 5 is more visible in the AIA 211 Å channel than in the other channels. The onset and progression of the Jet 5 minifilament eruption and the birth and growth of the spire are shown in Figure 2 by images selected from our AIA 211 Å movie (MOVIE 1), and can be seen with better time resolution in the movie itself. The third frame of Figure 2 is 2.0 minutes after the movie's first frame in which the minifilament can be **unambiguously** seen to be rising, the movie's frame taken at 11:33:00 UT. The rising and erupting minifilament is pointed out by the arrows in the third, fourth, fifth, and sixth frames of Figure 2. The seventh frame of Figure 2 is from the movie's first frame in which the spire is discernible. The arrow in the seventh frame points to the sprouting spire. The arrow in the eighth frame points to the rapidly growing spire 24 s after the seventh frame. In our judgment, the spire is most prominent in the 211 Å movie's frame from which the eleventh frame of Figure 2 was cropped, and which was taken four minutes after the seventh frame of Figure 2. In the twelfth frame of Figure 2, the spire is fading.

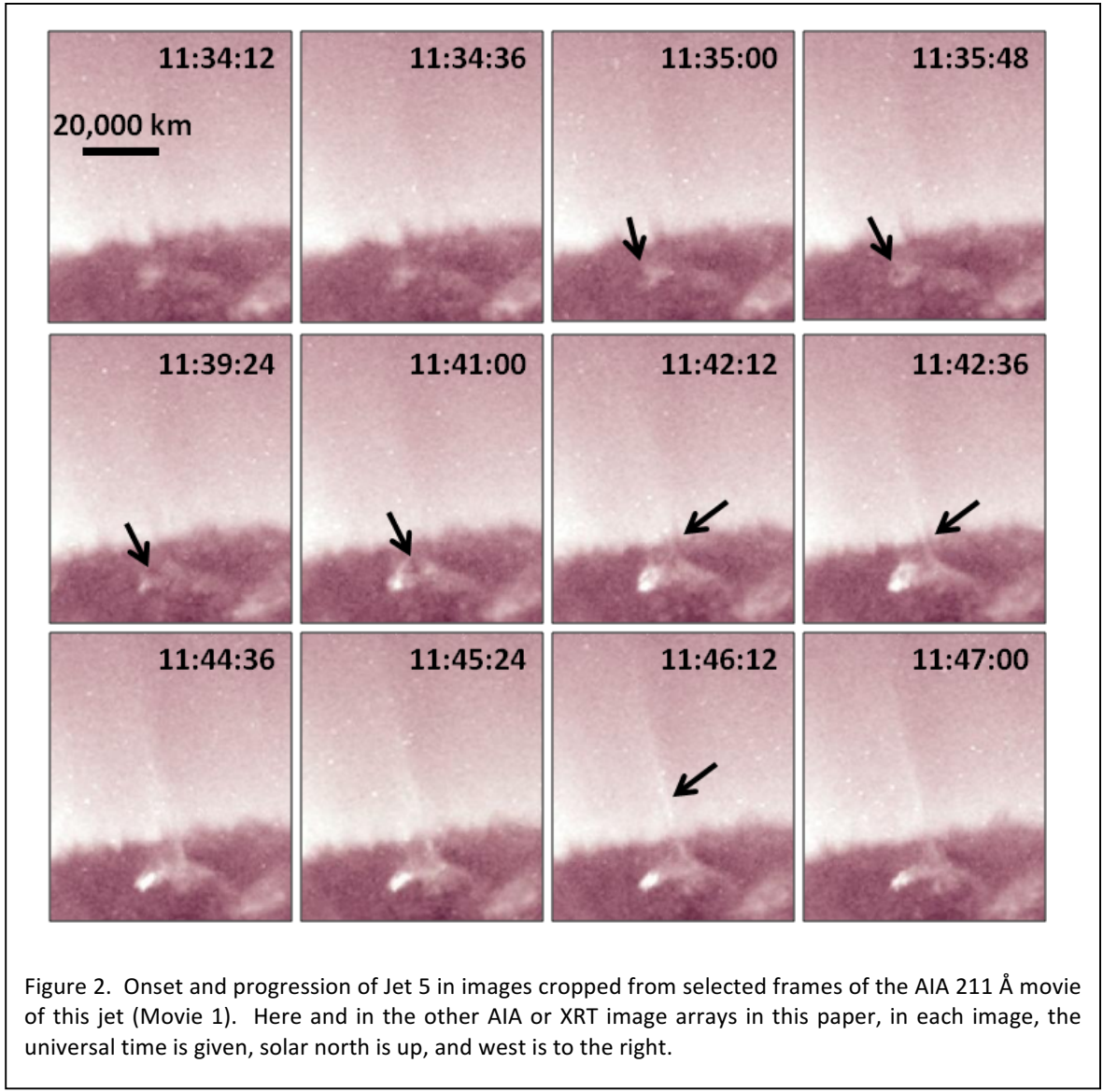

Figure 2. Onset and progression of Jet 5 in images cropped from selected frames of the AIA 211 Å movie of this jet (Movie 1). Here and in the other AIA or XRT image arrays in this paper, in each image, the universal time is given, solar north is up, and west is to the right.

In Figure 3, the onset and growth of Jet 5 are shown by images cropped from selected frames of the XRT base-difference movie (MOVIE 2). The corresponding coronal X-ray images cropped from the original (non-differenced) XRT movie (MOVIE 3) are in Figure 4. (For each of our three example jets, in the XRT base-difference movie and the XRT original movie provided as animations of the corresponding Figures, in the first frame of the movie an arrow points to the location at which the jet will occur.) Comparison of Figures 3 and 4 to Figure 2 shows that in both the base-difference XRT movie and the original XRT movie the JBP is as obvious as in the

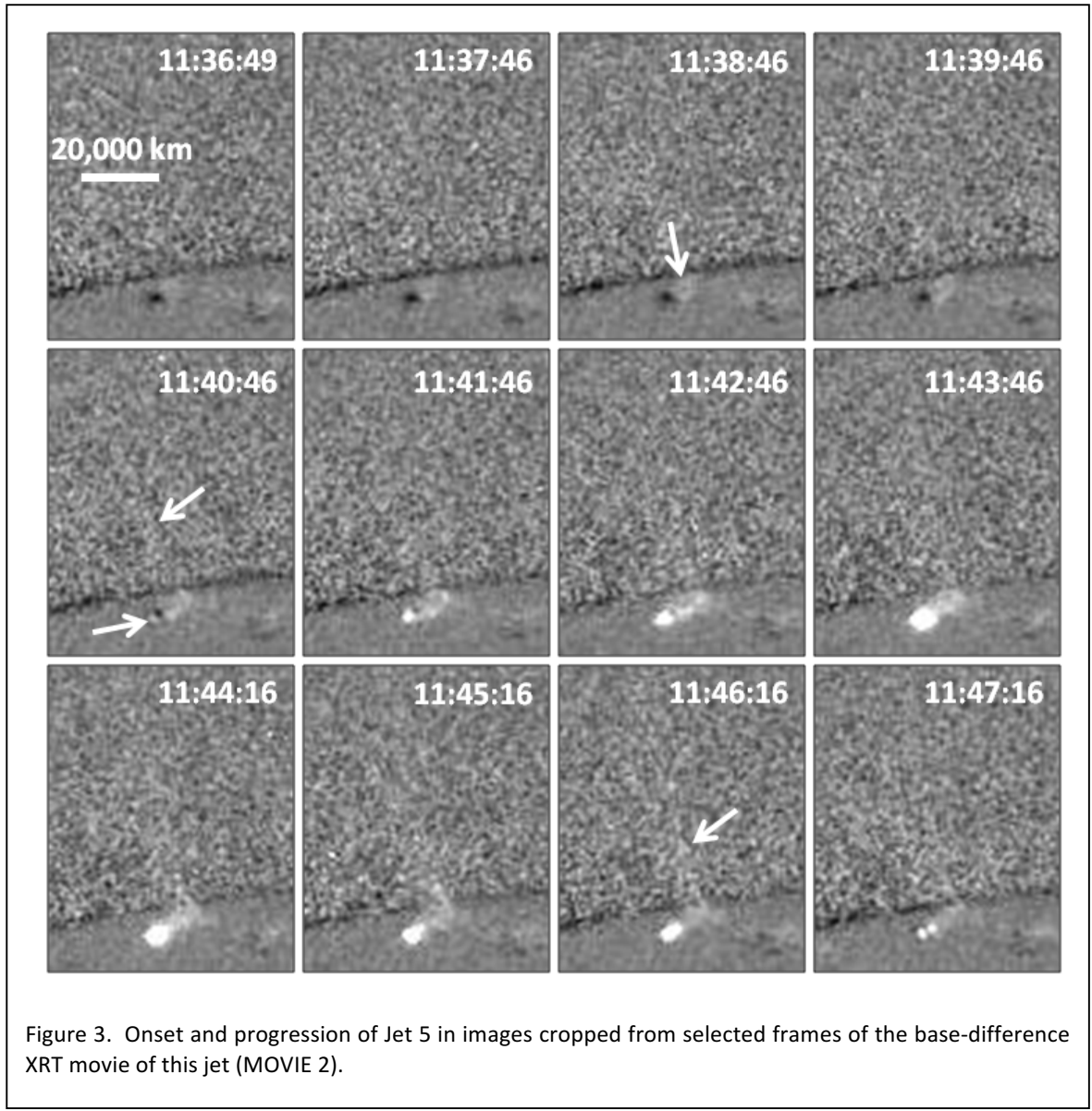

Figure 3. Onset and progression of Jet 5 in images cropped from selected frames of the base-difference XRT movie of this jet (MOVIE 2).

AIA 211 Å movie but the spire has less visibility in the XRT movies than in the AIA 211 Å movie. In MOVIE 2, the JBP and the spire both are first discernible at 11:40:46 UT, in the frame from which the fifth frame of Figure 3 was cropped. In that frame, the lower arrow points to the onset of the JBP and the upper arrow points to the first glimpse of the spire. The spire is more discernible in the base-difference XRT image than in the corresponding non-differenced XRT image in the fifth frame of Figure 4. In MOVIE 2, the onset of the JBP and spire in Jet 5 is seen to be preceded by the base interior brightening (BIB). The first frame in which sustained brightening in the base interior becomes discernible in MOVIE 2 is at 11:38:46 UT. The third frame of Figure 3 is from that frame of MOVIE 2, and the arrow in the third frame points to the onset of the BIB. Thus, in Jet 5, the onset of the minifilament's rise, at 11:33:00 UT in the AIA 211 Å movie, definitely preceded the onset of any discernible coronal X-ray brightening in the XRT movie. From stepping through both the base-difference XRT movie (MOVIE 2) and the non-differenced XRT movie (MOVIE 3), we judge that the spire is most prominent in the frames at 11:46:16 UT, from which the eleventh frames of Figures 3 and 4 were cropped. The arrow in the eleventh frames of Figures 3 and 4 points to the middle of the spire's extent above the limb. In the twelfth frames of Figures 3 and 4, the spire, the BIB, and the JBP are all fading out.

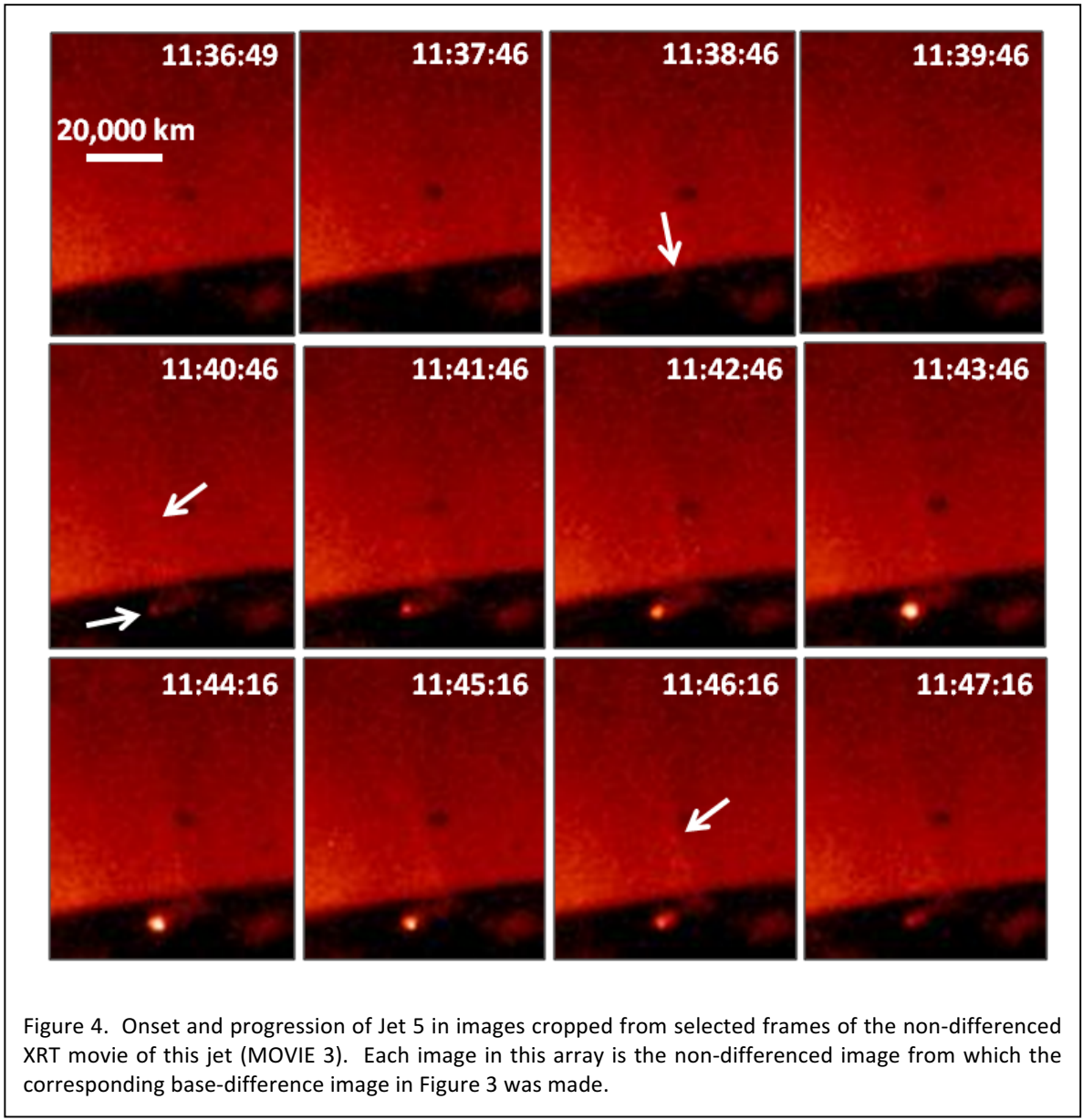

Figure 4. Onset and progression of Jet 5 in images cropped from selected frames of the non-differenced XRT movie of this jet (MOVIE 3). Each image in this array is the non-differenced image from which the corresponding base-difference image in Figure 3 was made.

*2.2.2. Jet 14*

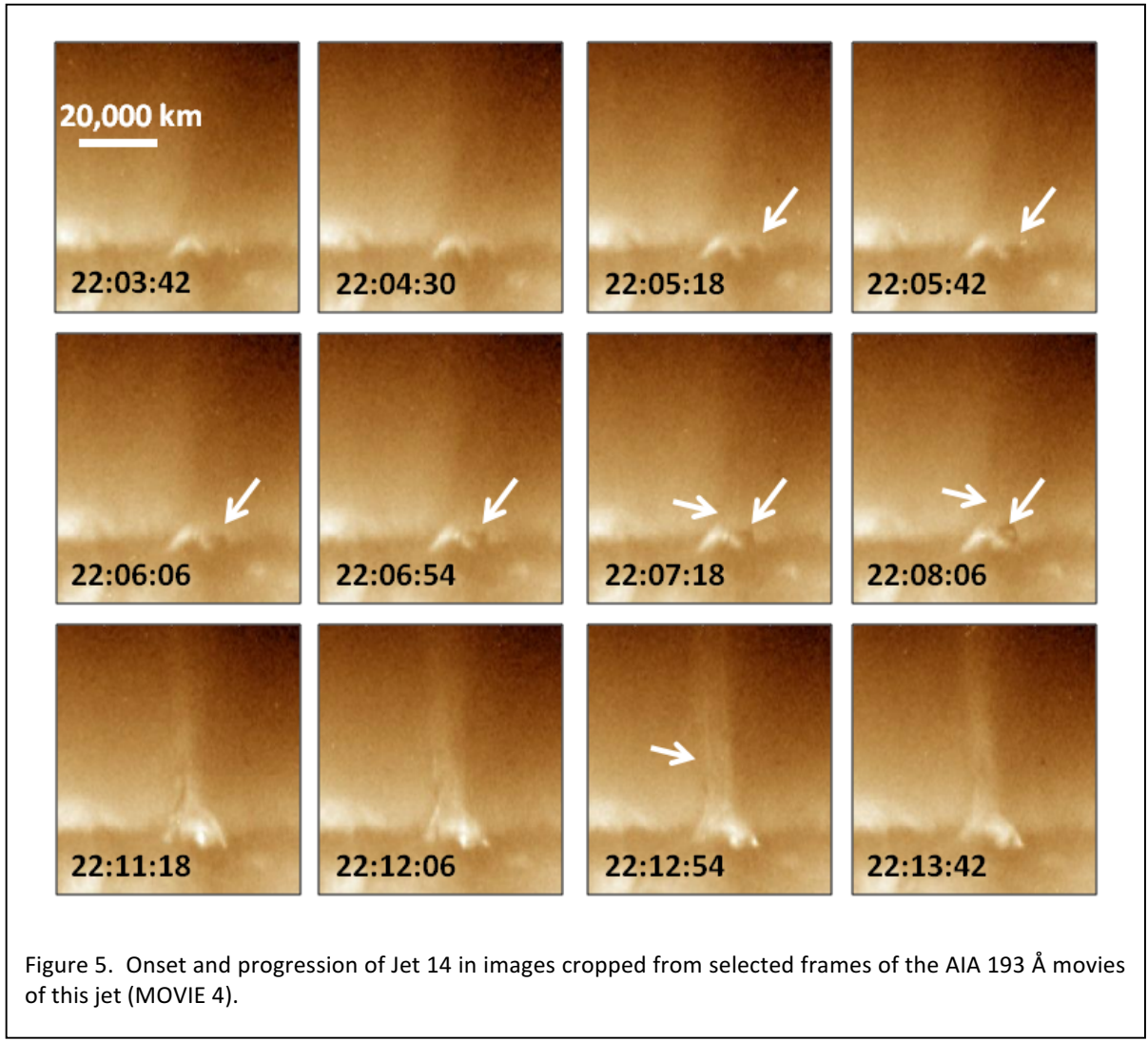

Figure 5. Onset and progression of Jet 14 in images cropped from selected frames of the AIA 193 Å movies of this jet (MOVIE 4).

The next example jet eruption is that of Jet 14 of Table 1. Jet 14 is a typical blowout jet in that in the XRT movie the spire is broader and brighter than in typical standard jets. In the AIA movie the spire is seen to have multiple-strand substructure. The eruption of this jet is an example of a jet eruption in which the minifilament rise in the AIA movie and the BIB in the XRT movie start simultaneously to within the time resolution of the two movies.

The erupting minifilament in Jet 14 is most visible in AIA's 193 Å band. Figure 5 shows the onset and progression of the minifilament eruption and the onset and progression of the jet spire in images cropped from the AIA 193 Å movie of Jet 14 (MOVIE 4). In the first two frames of Figure 5, the minifilament is barely discernible and is not yet discernibly rising. The third frame of Figure 5 is 1.2 minutes after the movie's first frame in which the minifilament is unambiguously seen to have started rising. Before that frame, the minifilament is embedded in the near-limb spicule forest, so that whether it has started rising cannot be unambiguously discerned. [In Jet 14 the minifilament is viewed end-on as in the cartoons in Figure 1. Sterling et al (2015) tailored those cartoons to fit this jet.] In the third-through-eighth frames of Figure 5, the arrows pointing down and to the left point to the erupting dark minifilament. In the ninth frame of Figure 5, the dark minifilament plasma has become a dark strand of the spire. The seventh frame of Figure 5 is from the movie's first frame in which the onset of the spire is faintly discernible. The arrows pointing to the right in the seventh and eighth frames of Figure 5 point to the faint growing spire. The eleventh frame of Figure 5 is from the movie's frame in which the spire is most prominent. In that frame, the spire has become about half as wide as the base and faintly shows multi-strand substructure. The arrow pointing to the spire in that frame is about a third of the way up along the spire's extent above the limb. In the twelfth frame of Figure 5, the spire has started to shrink in width and fade.

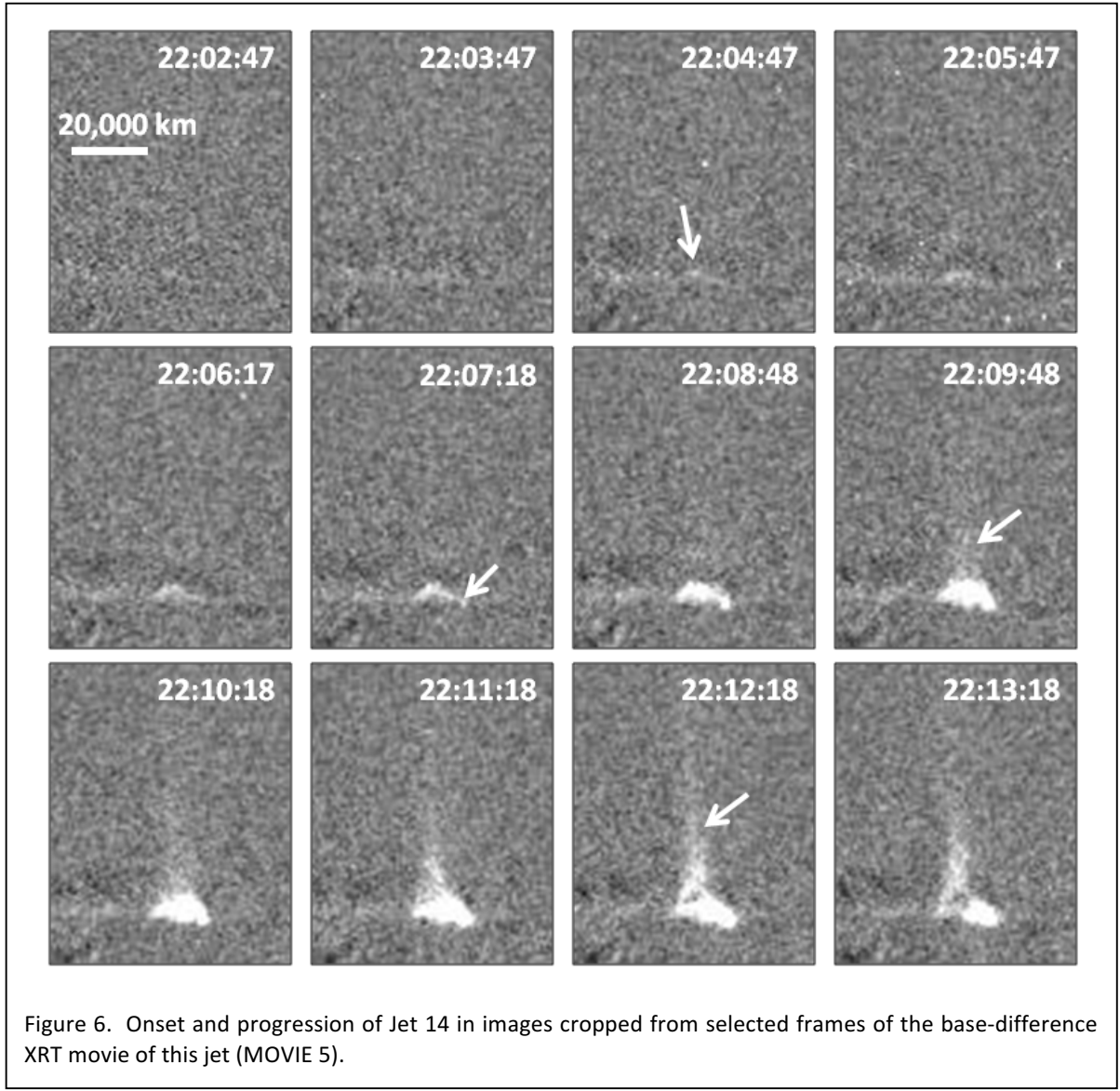

Figure 6. Onset and progression of Jet 14 in images cropped from selected frames of the base-difference XRT movie of this jet (MOVIE 5).

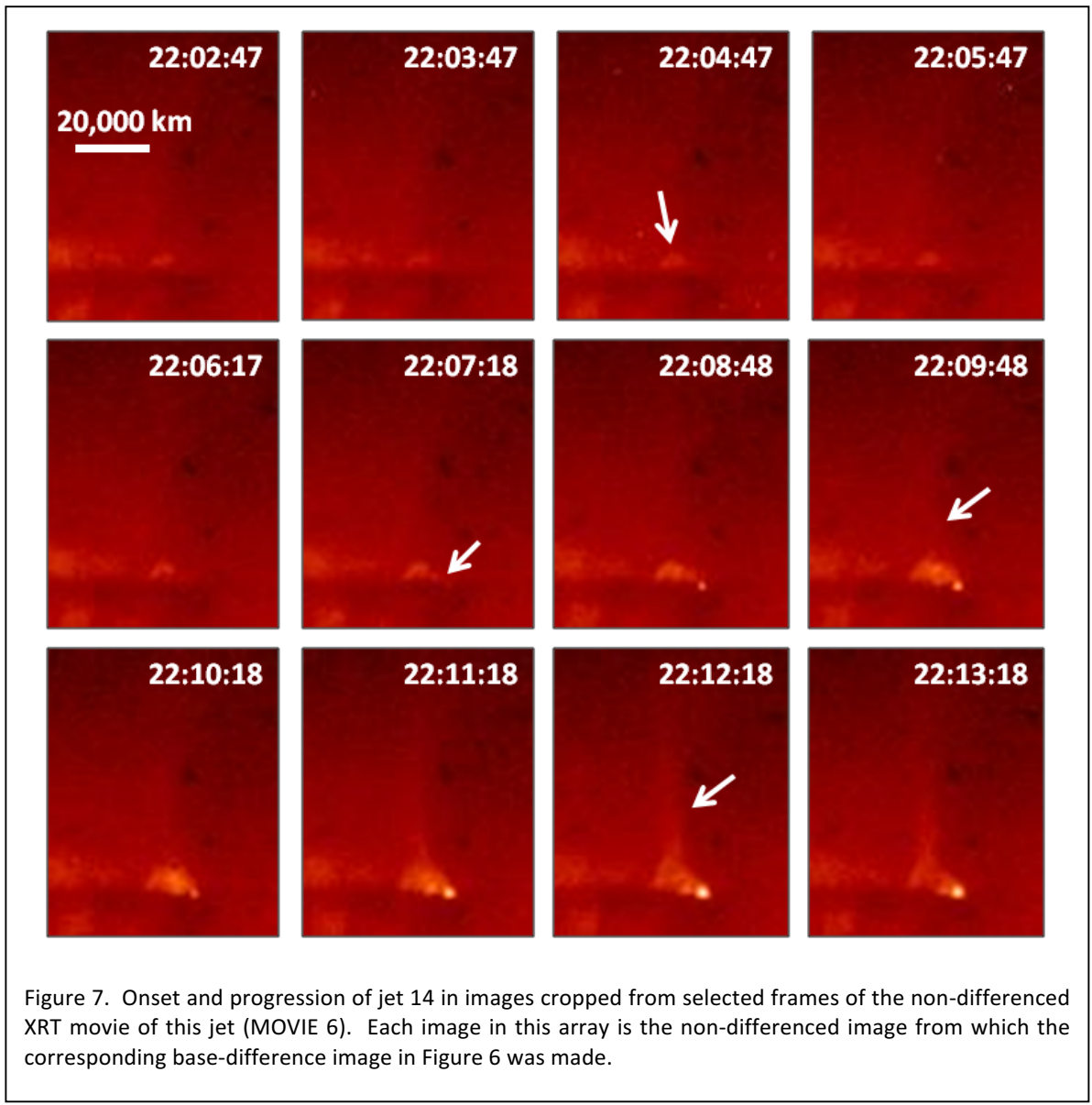

Figure 7. Onset and progression of jet 14 in images cropped from selected frames of the non-differenced XRT movie of this jet (MOVIE 6). Each image in this array is the non-differenced image from which the corresponding base-difference image in Figure 6 was made.

In Figure 6, the onset and progression of Jet 14 are shown in images cropped from selected frames of the XRT base-difference movie (MOVIE 5). The corresponding coronal X-ray images cropped from the original (non-differenced) XRT movie (MOVIE 6) are in Figure 7. As comparison of Figures 6 and 7 to Figures 3 and 4 shows, in the XRT movies the BIB and spire of Jet 14 (a blowout jet) are definitely brighter than the BIB and spire of Jet 5 (a standard jet), which is a typical difference of blowout jets from standard jets. Comparison of Figures 6 and 7 to Figure 5 shows that the spire of Jet 14 is somewhat less visible in the XRT movies than in the AIA movie, but the onset of the JBP is more visible in the XRT movies.

The third frame of Figure 6 is from the XRT base-difference movie's first frame in which the onset of the BIB is discernible. That brightening is also discernible in the original XRT movie, and is pointed to by the arrow in the third frames of Figures 6 and 7. In the first two frames of Figures **5** and 7, the base interior is seen as a bright mound, but there is not yet any discernible increase in the mound's brightness, as Figure 6 shows. Because our choice of the movie frame for the start of the minifilament's rise in the AIA movie and our choice of the frame for the start to the BIB in the XRT base-difference movie are each uncertain by ± 1 time step, the time range for the start of the minifilament's rise is from 22:03:42 UT to 22:04:30 UT and the time range for the start of the BIB is from 22:04:17 UT to 22:05:17 UT. Because these two time intervals overlap, the minifilament's rise and the BIB in Jet 14 start simultaneously within the uncertainty of the timing.

Figures 6 and 7 show that in the XRT movies the BIB in Jet 14 definitely starts before the JBP and the JBP definitely starts before the spire. The sixth frame of Figure 6 is from the XRT base-difference movie's first frame in which the onset of the JBP is first discernible. The corresponding image from the original XRT movie is the sixth frame of Figure 7. The arrow in the sixth frames of Figures 6 and 7 points to the JBP onset, which is not discernible by eye in that non-differenced XRT image in Figure 7 (but the JBP has become clearly visible in the next frame). The eighth frame of Figure 6 is from the XRT base-difference movie's first frame in which the spire is definitely discernible. The arrow in the eighth frame points to the growing spire. The eighth frames of Figures 6 and 7 together show that at this time the spire of Jet 14 is less visible in the original XRT movie than in the base-difference XRT movie. The eleventh frames of Figures 6 and 7 are from the movie's frame in which the jet spire is most prominent in the XRT base-difference movie. The arrow in the eleventh frames of Figures 6 and 7 points to the middle of the extent of the spire above the limb in the base-difference image.

### 2.2.3. Jet 7

Our third and last example jet eruption is that of Jet 7 of Table 1. Because Jet 7 happened inside the limb in the northern polar coronal hole, its base did not extend above the limb and is viewed against the disk in the AIA movie (MOVIE 7), in the base-difference XRT movie (MOVIE 8), and in the non-differenced orignial XRT movie (MOVEIE 9). Perhaps mainly because the spire is viewed against a moderately-bright diffuse background near and above the limb in the coronal hole, the spire is seen less clearly in Jet 7 than in typical blowout jets in which the spire is nearly all viewed above the limb, as in Jet 14. From the XRT movies, we judged that Jet 7 is a blowout jet because the growing spire shows two strands separated by a distance comparable to the width of the base of the jet. The eruption of jet 7 is an example of a jet eruption in which the BIB becomes discernible in coronal X-ray emission in the XRT movies before the rising of the minifilament is first unambiguously detected in the AIA movie. Faint nebulous upflow along the minifilament-side of the base arch is discernible in the AIA movie prior to the first detection of the BIB in the XRT movie (see Movie 7 and Table 1), suggesting that the minifilament's slow rise may be underway, but the rising minifilament cannot yet be seen the AIA movie because the minifilament is still hidden by the spicule forest, or because the minifilament's rising portion is very faint.

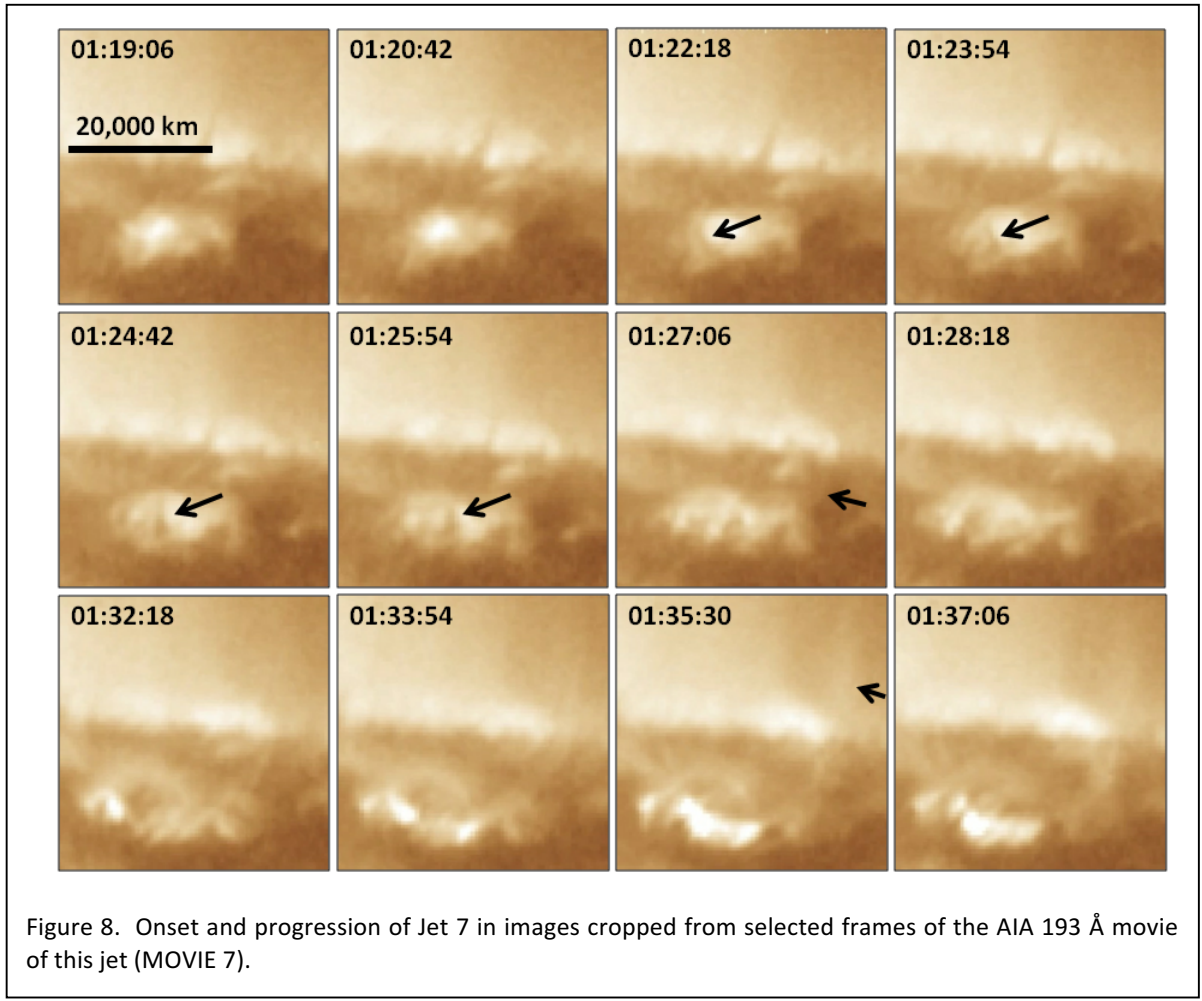

Figure 8. Onset and progression of Jet 7 in images cropped from selected frames of the AIA 193 Å movie of this jet (MOVIE 7).

Although the erupting minifilament and the spire in Jet 7 are more visible in AIA's 193 Å band than in AIA's other bands, these features are less distinctly visible in the AIA 193 Å movie of Jet 7 (MOVIE 7) than their corresponding features in either the AIA 211 Å movie of Jet 5 (MOVIE 1) or the AIA 193 Å movie of Jet 14 (MOVIE 4). Figure 8 shows the onset and progression of the minifilament eruption and the onset and progression of the spire in Jet 7 in AIA 193 Å images cropped from MOVIE 7. Only the stronger of the spire's two strands discernible in the XRT movies of Jet 7 becomes discernible in the AIA 193 Å movie, MOVIE 7. In MOVIE 7, in accord with Jet 7 being a blowout jet, it appears that the western leg of the erupting minifilament becomes that strand as that strand becomes discernible.

In the first two frames of Figure 8, the minifilament is barely visible and is not discernibly rising. The third frame of Figure 8 is from the movie's first frame in which the minifilament has discernibly moved upward. The arrows in the third-through-sixth frames point to the rising minifilament. The seventh frame of Figure 8 is from the movie's first frame in which the spire's stronger strand has become faintly discernible. The arrow in the seventh frame points to the first glimpse of the growing strand. The eleventh frame of Figure 8 is from the AIA movie's frame in which we judge that strand of the spire to be most prominent, mainly in having its greatest discernible extent (reaching the top of the frame). The arrow in the eleventh frame points to the middle of that strand's extent above the base of the jet.

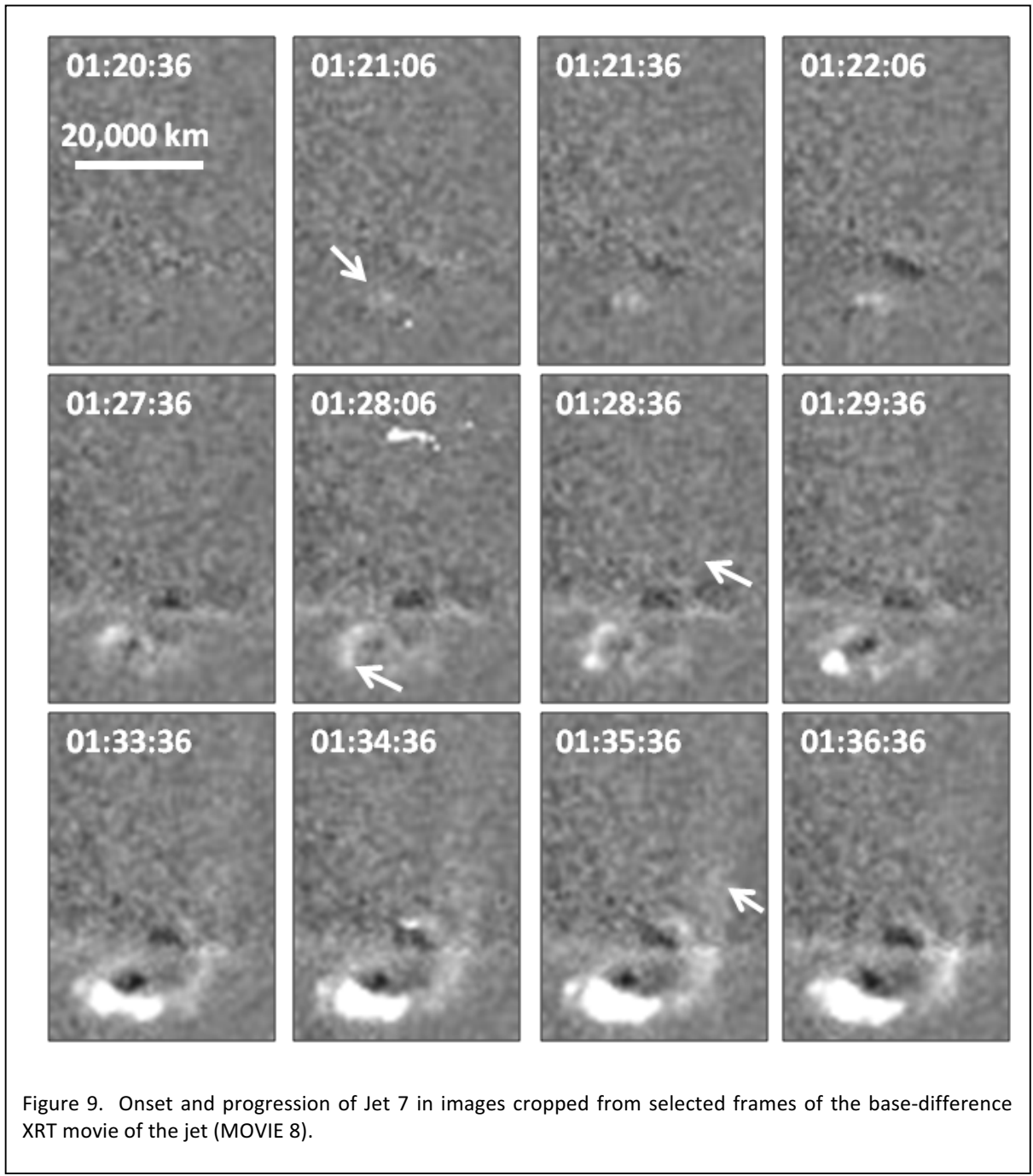

Figure 9. Onset and progression of Jet 7 in images cropped from selected frames of the base-difference XRT movie of the jet (MOVIE 8).

Figure 9 shows the onset and progression of Jet 7 in images cropped from selected frames of the XRT base-difference movie (MOVIE 8), and Figure 10 has the corresponding coronal X-ray images cropped from the non-differenced XRT movie (MOVIE 9). The second frames of Figures 9 and 10 are from the XRT movie's first frame in which the start of the BIB in Jet 7 is discernible in the XRT base-difference movie. The arrow in the second frames points to the BIB's onset.

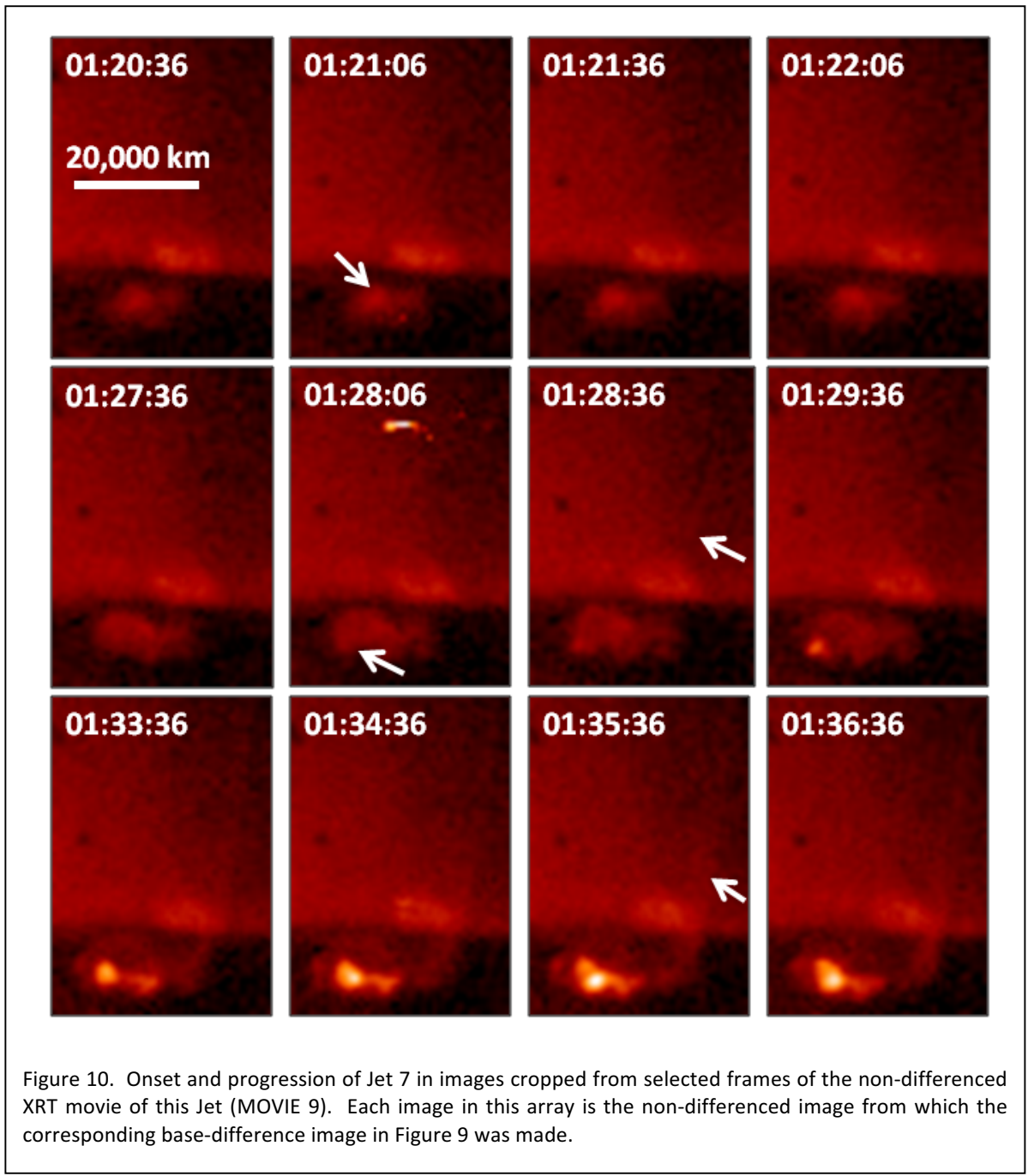

Figure 10. Onset and progression of Jet 7 in images cropped from selected frames of the non-differenced XRT movie of this Jet (MOVIE 9). Each image in this array is the non-differenced image from which the corresponding base-difference image in Figure 9 was made.

In the XRT base-difference movie, the spire is composed of two separate strands which turn on and evolve together. From their start at about 01:28 UT, the eastern strand is the fainter of the two. During the next five minutes, the eastern strand fades out while the western strand broadens to become the entire spire in the spire's maximum. The seventh frames of Figures 9 and 10 are from the XRT movie's first frame in which the spire is faintly discernible in the XRT base-difference movie. The arrow in the seventh frames points to the western strand; the eastern strand is just barely discernible in the seventh-frame base-difference image. Figures 9 and 10 together with Figure 8 show that in Jet 7 the BIB – which presumably brightened from heating by external reconnection of the magnetic arcade having the minifilament in its core – had started by 01:21:06 UT, roughly a minute before the minifilament's rise was first **unambiguously** detected in the AIA 193 Å movie.

The sixth frames of Figures 9 and 10 are from the XRT movie's first frame in which the start of the JBP in Jet 7 is discernible in the XRT base-difference movie. The JBP obviously grows brighter and larger in the seventh-through-ninth frames.

In the final four frames of Figure 9, the eastern strand is no longer discernible as the western strand evolves through its maximum. The eleventh frames of Figures 9 and 10 are from the frame of the two XRT movies in which we judge the spire to be most prominent. The arrow in the eleventh frames points to the middle of the spire's extent above the limb.

## 3. RESULTS AND INTERPRETATION

### 3.1. Observed Sequence of Events in Our Jets

In this Section we examine whether the event sequences among our 15 jets are similar to corresponding event sequences among flare eruptions that produce a CME. We do that by examining, for each jet individually and for all 15 jets in aggregate, the relative timing of the seven events that we observed in each jet, the seven events for which the times are given in Table 1: (1) the start of the minifilament rise in the AIA movie, (2) the start of the JBP in the XRT movie, (3) the start of the BIB in the XRT movie, (4) the start of the spire in the XRT movie, (5) the start of the spire in the AIA movie, (6) the spire maximum in the XRT movie, and (7) the spire maximum in the AIA movie.

#### 3.1.1. Time Lapse from Start of BIB to Start of Spire

Figure 11 is a histogram of the time lapse from the start of the BIB in the XRT movie to the start of the spire in the XRT movie for our 15 jets. Each elementary block of this histogram is for the jet whose number is the number in that block; the BIB-to-spire time lapse in that jet falls in the 1-minute time-difference interval of that block. If a jet's time lapse fell on the line between two blocks, the jet was assigned to the block sitting farther to the right on the abscissa. The time lapse is positive (negative) if the BIB started before (after) the spire started. Corresponding to the histogram in Figure 11, Figure 12 is the histogram of the time lapse from the start of the BIB in the XRT movie to the start of the spire in the AIA movie. These two histograms together show that the BIB in the XRT movie nearly always started before the spire did in either the XRT movie or the AIA movie. In only Jets 11 and 15 did the XRT BIB possibly start after the XRT spire (within the ~ ±1 minute uncertainty in the time lapse obtained from Table 1), and in only Jets 12 and 15 did the XRT BIB possibly start after the AIA spire.

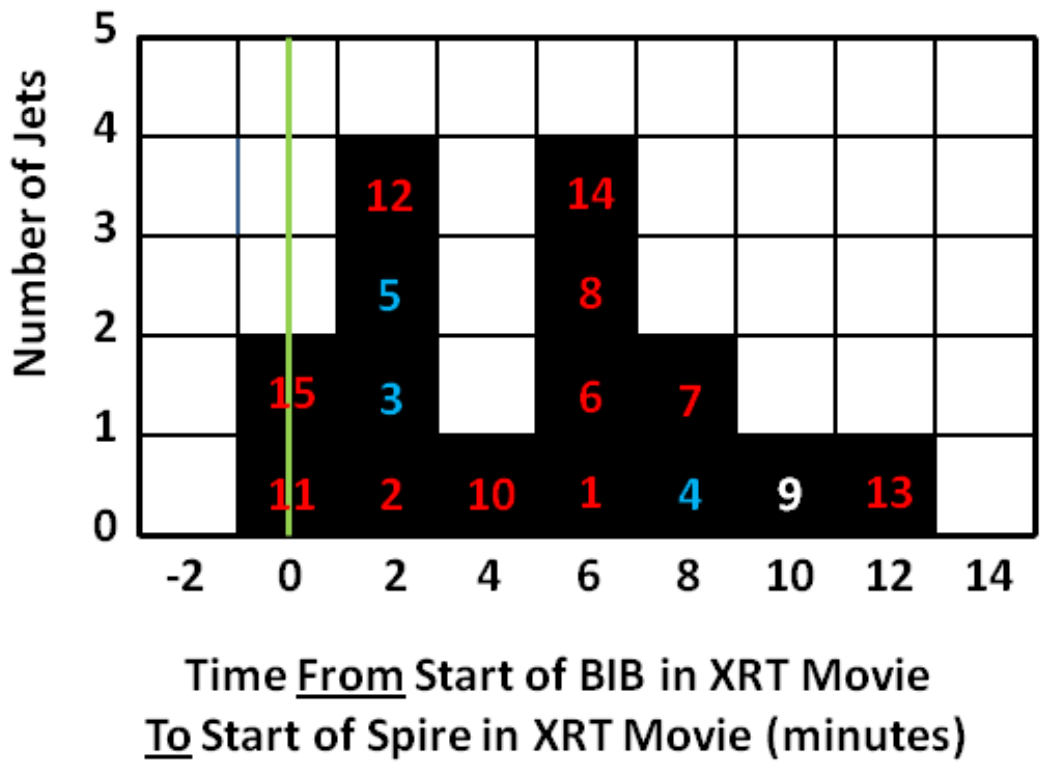

Figure 11. Histogram of time lapse from start of BIB in XRT movie to start of spire in XRT movie, for our 15 jets. In this histogram and the other histograms in this paper, the number in each elementary block is the number of the jet in Table 1 that falls in that block; the numbers are red for the blowout jets and blue for the standard jets, and number 9 is white, denoting that the morphology of Jet 9 is an ambiguous mix of standard-jet morphology and blowout-jet morphology.

The mean time lapse of the Figure-12 histogram is marginally longer than the mean time lapse of the Figure-11 histogram. That difference results from what is shown by the histogram in Figure 13, the histogram of the time lapse from the start of the XRT spire to the start of the AIA spire. The Figure-13 histogram shows that in twelve of the jets the spire started at nearly the same time (within about 2 minutes of each other) in the XRT movie as in the AIA movie, but in the other three jets (Jets 3, 8, and 13) the spire definitely started later in the AIA movie than in the XRT movie.

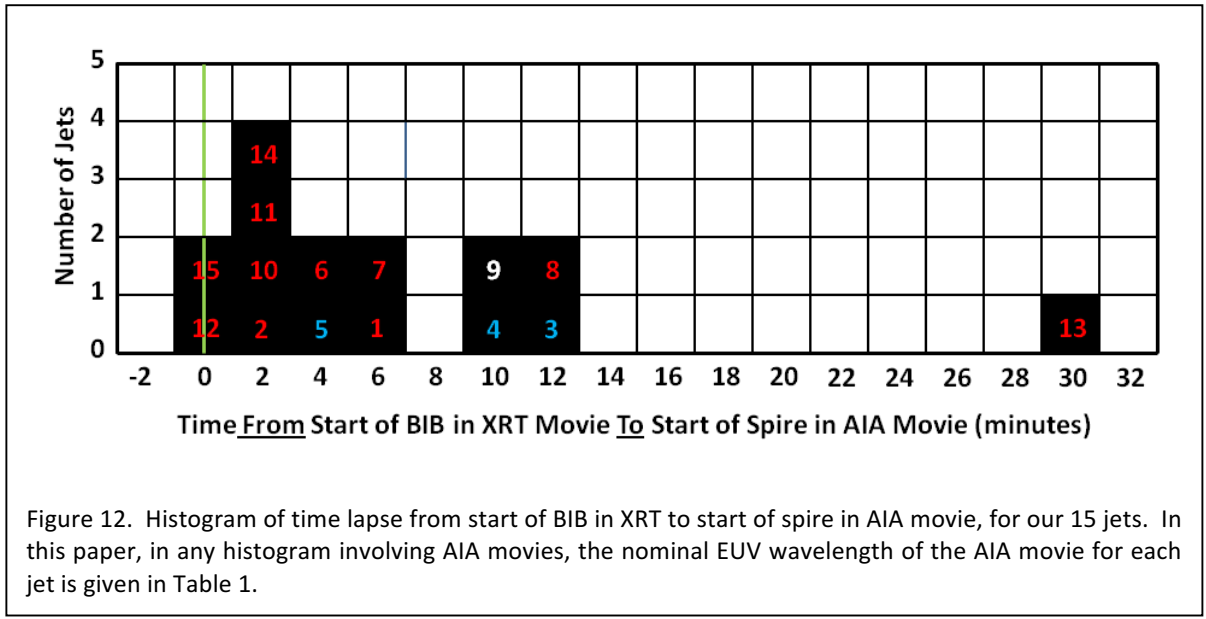

Figure 12. Histogram of time lapse from start of BIB in XRT to start of spire in AIA movie, for our 15 jets. In this paper, in any histogram involving AIA movies, the nominal EUV wavelength of the AIA movie for each jet is given in Table 1.

In the Sterling et al (2015) scenario and in all other jet-production scenarios in which the spire is produced by external reconnection of closed field in the base of the jet with ambient open field, the reconnection simultaneously produces both new open field lines and new closed field loops, both of which are rooted in the base of the jet. Reconnection-heated plasma in the new open field flows out along that field as the spire, and reconnection-heated plasma also fills the new closed loops. So, one might expect that the spire and the BIB would become discernible nearly simultaneously in the XRT movie. Instead, in our 15 jets, the XRT BIB was nearly always discernible before both the XRT spire and the AIA spire. This is perhaps not surprising because the heated plasma on the new open field readily expands, decreasing its density and hence its emissivity of thermal X-ray and EUV radiation, whereas the heated plasma in the new closed loops is trapped in them and compressed to higher density as the new closed loops contract downward from the reconnection site. In any case, in each of our 15 jets the onset of the XRT BIB is either the first sign or the tied-for-first sign that the external reconnection has started.

### 3.1.2. Sequence of Events in the Eruption Onset

Table 2 lists for each of our 15 jets the observed sequence of three events (i.e., first detections) early in the eruption: the start of the minifilament rise in the AIA movie (second column of Table 2), the start of the JBP in the XRT movie (third column), and the start of the BIB in the XRT movie (fourth column). For each jet, the entry for that jet in each event column is the order of the occurrence of that event relative to the other two events in the jet's eruption. If two of the three events occurred simultaneously within the uncertainty of their times of occurrence, that is, if the uncertainty time ranges of the two events overlapped, then both events are given the same place in order of occurrence. That happened in three jets: Jets 1, 12, and 15. In Jets 1 and 12, First Detection of Minifilament Rise in AIA and First Detection of BIB in XRT tied for first, and First Detection of JBP in XRT was third. In Jet 15, First Detection of Minifilament Rise in AIA was first and the other two events tied for second place. Table 2 also shows that in none of our 15 jets did the uncertainty time range of each event overlap with the uncertainty time range of each of the other two events. That is, in none of our jet eruptions did all three events tie for first place.

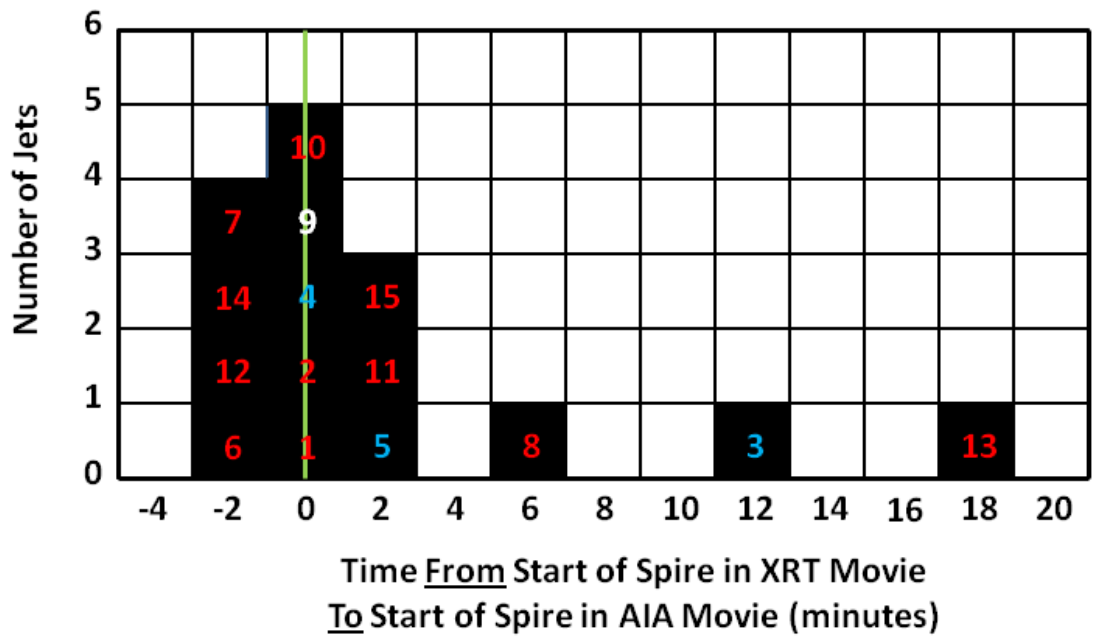

Figure 13. Histogram of time lapse from start of spire in XRT movie to start of spire in AIA movie, for our 15 jets.

In the Sterling et al (2015) minifilament-eruption schematic for jet production (Figure 1), the order of the three event columns from left to right in Table 2 is the order expected if minifilament rise starts (by some ideal MHD process) before the runaway internal reconnection that makes the JBP, and the runaway internal reconnection starts before the runaway external reconnection that makes the BIB. That is, for that sequence of events in the jet eruption, we would expect the entries in the three event columns in Table 2 to be, from left to right: First, Second, Third. No jet in Table 2 definitely shows evidence of that ordering of events. In only two jets (Jet 11 and Jet 15) is the observed sequence of events (sequence of first detections) possibly consistent with that ordering. For Jet 11, First Detection of BIB in XRT is third and First Detection of JBP in XRT is possibly second; for Jet 15, First Detection of Minifilament Rise in AIA is first and First Detection of BIB in XRT is possibly third. In the other 13 jets in Table 2, the first

| Jet Number | Table 2. Sequences of First Detections of Rising Minifilament, Jet-Base Bright Point, and Base Interior Brightening | | |
|---|---|---|---|
| | Event | | |
| | First Detection of Rising Minifilament (AIA) | First Detection of Jet-Base Bright Point (XRT) | First Detection of Base Interior Brightening (XRT) |
| 1 | First | Third | First |
| 2 | First | Third | Second |
| 3 | First | Third | Second |
| 4 | First | Third | Second |
| 5 | First | Third | Second |
| 6 | First | Third | Second |
| 7 | Second* | Third | First |
| 8 | First | Third | Second |
| 9 | Second* | Third | First |
| 10 | First | Third | Second |
| 11 | First | First | Third |
| 12 | First | Third | First |
| 13 | Second* | Third | First |
| 14 | First | Third | First |
| 15 | First | Second | Second |

* In the AIA movie of this jet, nebulous upflow from where the rising filament later appears is faintly discernible up to several minutes before the time of First Detection of Base Interior Brightening in the XRT movie (see Table 1), suggesting that the breakout reconnection that made the base interior brightening may have started concurrent with or after the start of the minifilament's slow rise but before the minifilament had risen enough to be no longer hidden in the spicule forest in the AIA movie.

detection of the JBP lags the first detections of both the rising minifilament and the BIB. Evidently, in roughly 85% (13/15) of polar X-ray jets, the runaway internal reconnection starts after the rise of the minifilament starts and lags the start of the runaway external reconnection. In eight of our jets, Jets 2, 3, 4, 5, 6, 8, 10, and 15, the minifilament started rising before the first sign of the start of either the external (breakout) reconnection or the internal reconnection. So, in these eight jet-eruption onsets, our observations allow the eruption to have started by an ideal MHD instability. In six of our jets, Jets 1, 7, 9, 12, 13, and 14, the first sign of breakout reconnection occurred either before or simultaneously with the First Detection of Minifilament Rise in AIA. So, in these six jet-eruption onsets, our observations allow the eruption to have been initiated by breakout reconnection. In the only other of our 15 jets, Jet 11, First Detection of Minifilament Rise in AIA and First Detection of JBP in XRT were simultaneous to within the uncertainty of their times and preceded First Detection of BIB in XRT. So, in this jet eruption, our observations allow the eruption to have started by runaway internal tether-cutting reconnection. Thus, our observations indicate that the sequences of occurrence of start of minifilament eruption, start of breakout reconnection, and start of runaway internal tether-cutting reconnection, in the eruption onsets of polar coronal X-ray jets have the same diversity as the sequences of corresponding events in the onsets of the much larger magnetic eruptions that make a flare and CME and have access to breakout reconnection. This result from Table 2 supports the picture of Sterling et al (2015) that the minifilament eruptions that make a polar coronal X-ray jet are miniatures of the filament eruptions that make a flare and CME and have access to breakout reconnection.

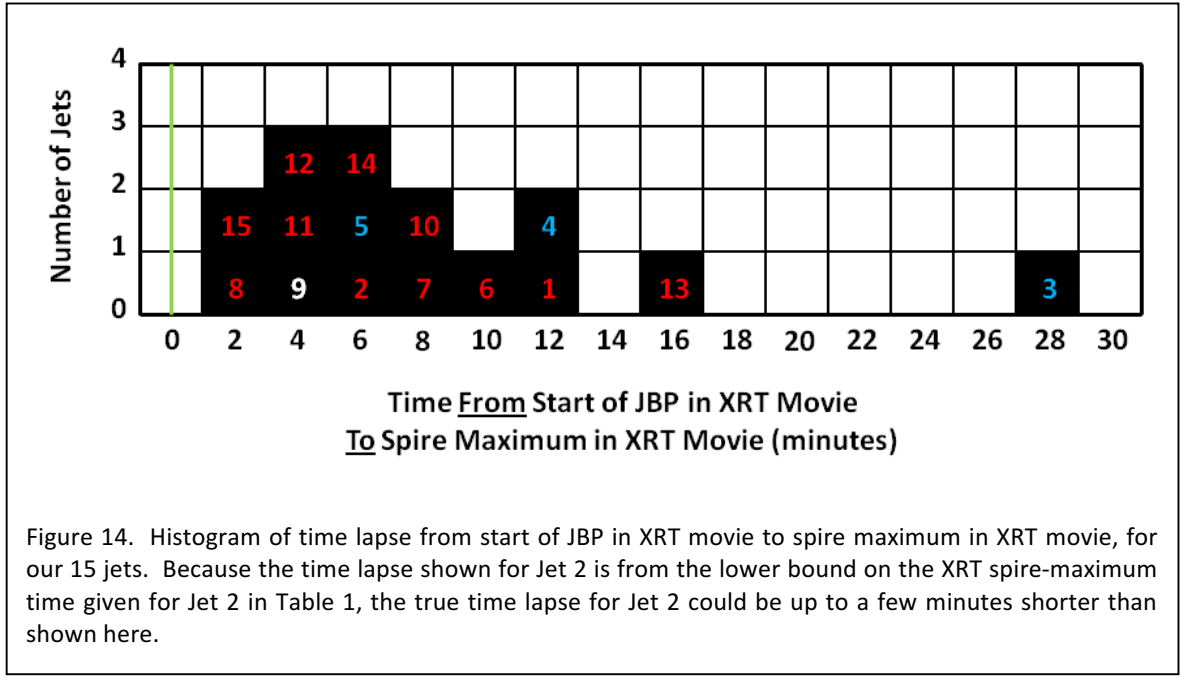

Figure 14. Histogram of time lapse from start of JBP in XRT movie to spire maximum in XRT movie, for our 15 jets. Because the time lapse shown for Jet 2 is from the lower bound on the XRT spire-maximum time given for Jet 2 in Table 1, the true time lapse for Jet 2 could be up to a few minutes shorter than shown here.

### 3.1.3. Time Lapse from Start of JBP to Spire Maximum

It is observed that in filament eruptions that make a flare and CME, the filament and the magnetic arcade that carries the filament in its sheared-field core erupt together in synchrony. The eruption typically begins with a slow rise of the filament and its enveloping arcade and then transitions to a much faster rise. The transition from slow rise to fast rise occurs during the onset of the impulsive phase of the flare, during which there is an impulsive burst of nonthermal hard X-ray emission and a steep rise in thermal X-ray emission (e.g., Zhang et al 2001, 2004; Sterling & Moore 2004b, 2005). Those emissions come from the feet and body of the flare arcade, which is built and heated by the runaway tether-cutting reconnection that occurs below the erupting filament-carrying flux rope, the core of the erupting arcade. As the reconnection builds and heats the flare arade, it also grows and heats the rising flux rope, and progressively cuts the growing flux rope's magnetic ties to the photosphere, which results in the flux rope rising faster (e.g., Moore et al 2001).  So, the impulsive-phase flare emission and the fast-rise phase of the eruption plausibly both result from a burst in the rate of tether-cutting reconnection.  This expected role of the runaway tether-cutting reconnection in producing the explosive fast phase of the eruption has been demonstrated in MHD simulations of magnetic eruptions that produce a flare and CME (e.g., Karpen et al 2012).

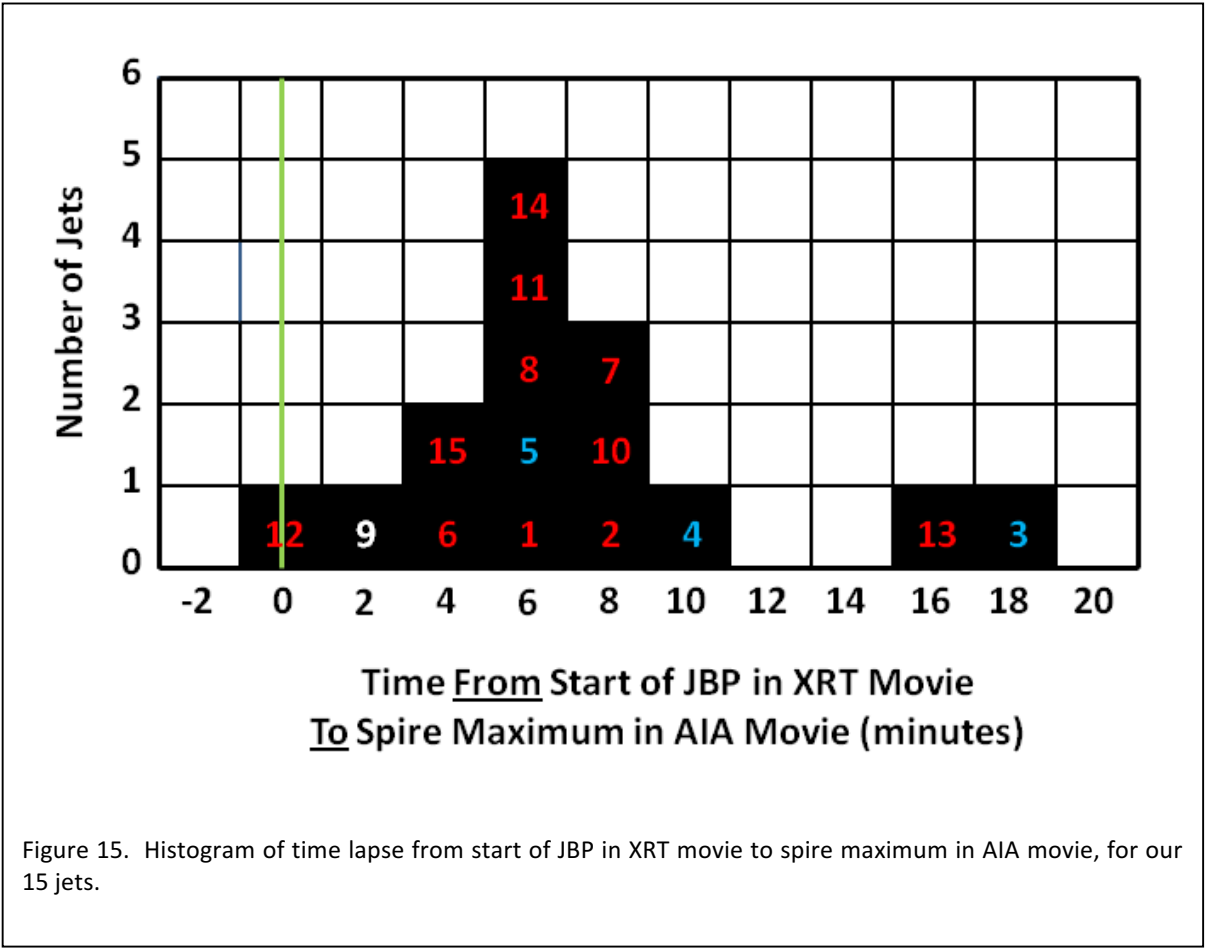

Figure 15. Histogram of time lapse from start of JBP in XRT movie to spire maximum in AIA movie, for our 15 jets.

   If the minifilament eruptions that make a polar coronal X-ray jet are miniature versions of the filament eruptions that make a flare and CME, then we should expect them to have the following two characteristics.  First, the eruption of the jet-base-edge arcade that carries the minifilament in its core should begin with a slow rise and then become much faster and more explosive.  This is observed to be a characteristic of the minifilament eruptions in our 15 jets (Sterling et al 2015).  Second, the explosive phase of the eruption should start after the internal runaway tether-cutting reconnection that builds and heats the JBP has started.  That should result in the breakout reconnection of the erupting arcade with the ambient open field (the reconnection that produces the spire) being much more strongly driven after the start of the JBP than before.  So, if the magnetic eruptions that make a polar X-ray jet work like the magnetic eruptions that make a flare and CME, we should expect the JBP to start before spire maximum occurs. In this Section, we show that in all of our 15 jets, the JBP evidently starts before spire maximum.

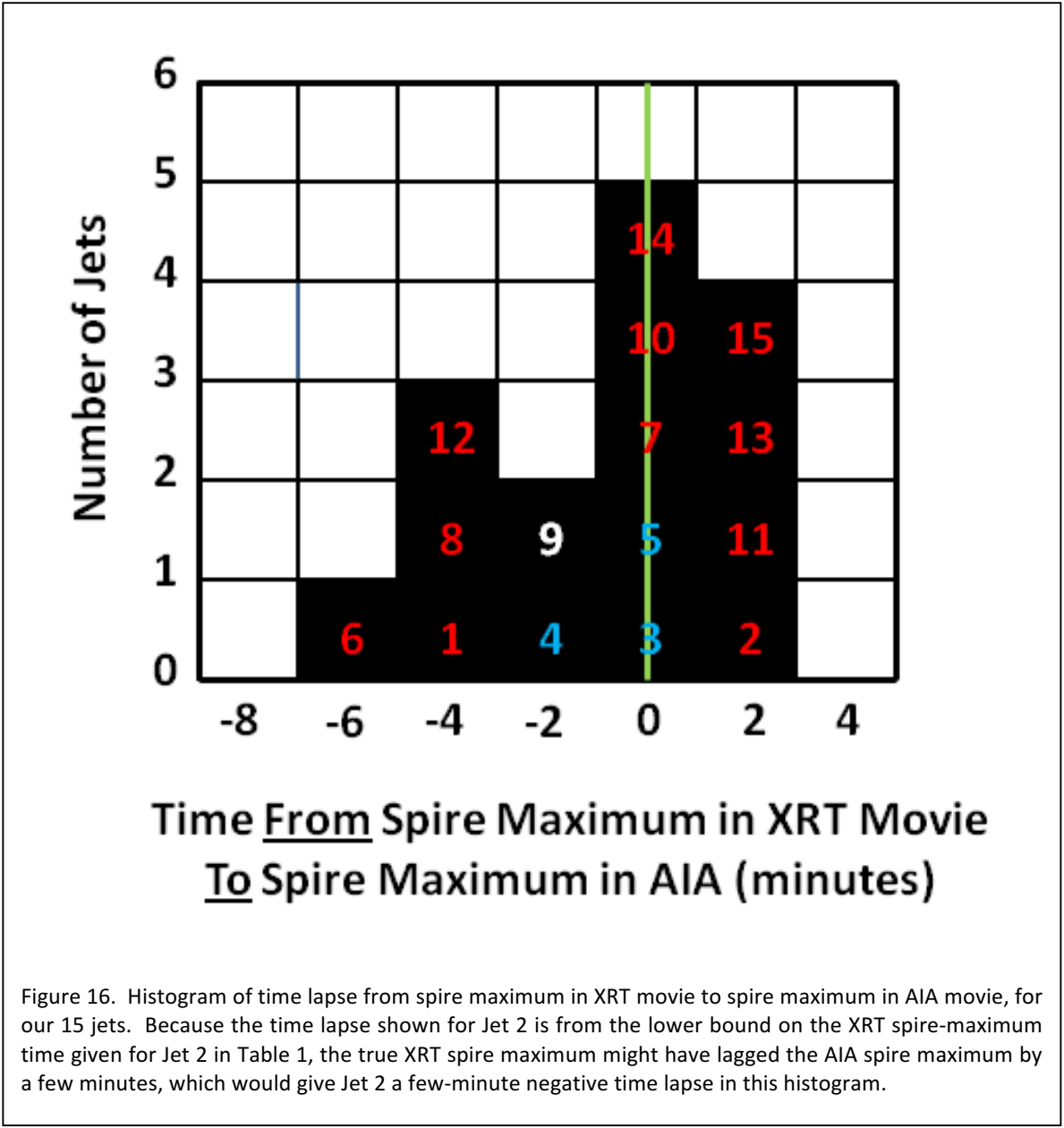

Figure 16. Histogram of time lapse from spire maximum in XRT movie to spire maximum in AIA movie, for our 15 jets. Because the time lapse shown for Jet 2 is from the lower bound on the XRT spire-maximum time given for Jet 2 in Table 1, the true XRT spire maximum might have lagged the AIA spire maximum by a few minutes, which would give Jet 2 a few-minute negative time lapse in this histogram.

For our 15 jets, Figure 14 is the histogram of the time lapse from the start of the JBP in the XRT movie to the spire maximum in the XRT movie. Figure 15 is the corresponding histogram of the time lapse from the start of the JBP in the XRT movie to the spire maximum in the AIA movie. The Figure-14 histogram shows that in the XRT movie of each of our jets, the JBP starts before the spire attains its maximum. The Figure-15 histogram shows that, in every jet except one (Jet 12), the start of the JBP in the XRT movie also leads the spire maximum in the AIA movie. For our 15 jets, Figure 16 is the histogram of the time lapse from spire maximum in the XRT movie to spire maximum in the AIA movie. It shows that in our set of jets there is some tendency for the AIA spire maximum to lead the XRT spire maximum, and that in Jet 12 the AIA spire maximum leads the XRT spire maximum by about 4 minutes. Through both the AIA spire maximum and the XRT spire maximum, the spire in Jet 12 stands out strongly in the XRT movie, but is only faintly visible in our AIA movie for Jet 12, which is from the AIA 193 Å channel. This suggests that through spire maximum in both AIA and XRT most of the plasma in the spire of Jet 12 was too hot to emit strongly in the AIA 193 Å channel, and hence that the XRT movie shows the "true" spire maximum in Jet 12 better that the AIA 193 Å movie. In any case, the histograms in Figures 14-16 are consistent with the explosive magnetic eruption that drives a polar X-ray jet being a miniature version of the explosive magnetic eruption that drives a flare and CME.

### 3.2. Explosive Lobe's Extent at the Start of Its Eruption

Table 2 lists the observed order of occurrence of three events in the onset and growth of the eruption in each of our 15 jets: the first detection of the rising minifilament, the first detection of the JBP (runaway internal reconnection), and the first detection of the BIB (runaway external reconnection). The Sterling et al (2015) scenario (depicted in Figure 1) for jet production by minifilament eruption allows each of the observed event sequences listed in Table 2. Sterling et al (2015) however did not explicitly address the question of the initial outer extent of the explosive magnetic lobe, the lobe whose core is the minifilament flux rope. In this Section, we consider how the order in which the three events of Table 2 happen in an observed jet reflects the explosive lobe's extent when it starts erupting. That is, we consider how the event order that occurs reflects how much of the void above the minifilament flux rope in the first drawing in Figure 1 is filled by the explosive lobe at the start of the slow rise of its eruption.

In the first drawing in Figure 1, only the core of the explosive lobe is drawn: the curled field line holding the cool-plasma minifilament represents the twisted-flux-rope core of the explosive lobe. The rest of the explosive lobe, its outer part, is not shown but is assumed to occupy some or all of the void, the region that is left empty of drawn field lines, above the flux rope.

If the outer envelope of the explosive lobe is only a thin shell of closed field loops closely encasing the flux rope, then much of the void is not filled by the explosive lobe, leaving room for the possibility of a null point in the magnetic field between the explosive lobe's outer envelope and the oppositely-directed open field on the far side of the void (Figure 1, first drawing). In this case, runaway external reconnection plausibly does not start simultaneously with the start of the eruption (the start of the slow rise of the minifilament) but somewhat later, when the explosive lobe has erupted far enough up along the outside of the stable lobe to compress the null region into a current sheet at which the external reconnection can start. That is, when the explosive lobe initially does not fill much of the void, the eruption plausibly starts either by ideal MHD instability without any initial runaway internal reconnection or by runaway internal reconnection, but not by runaway external reconnection. So, in Table 2, in any jet in which the base interior brightening was not detected first or tied for first, the explosive lobe is allowed to fill only part of the void at the start of the lobe's eruption (that is, a magnetic null is allowed in the void). That is allowed in nine jets (Jets 2, 3, 4, 5, 6, 8, 10, 11, 15).

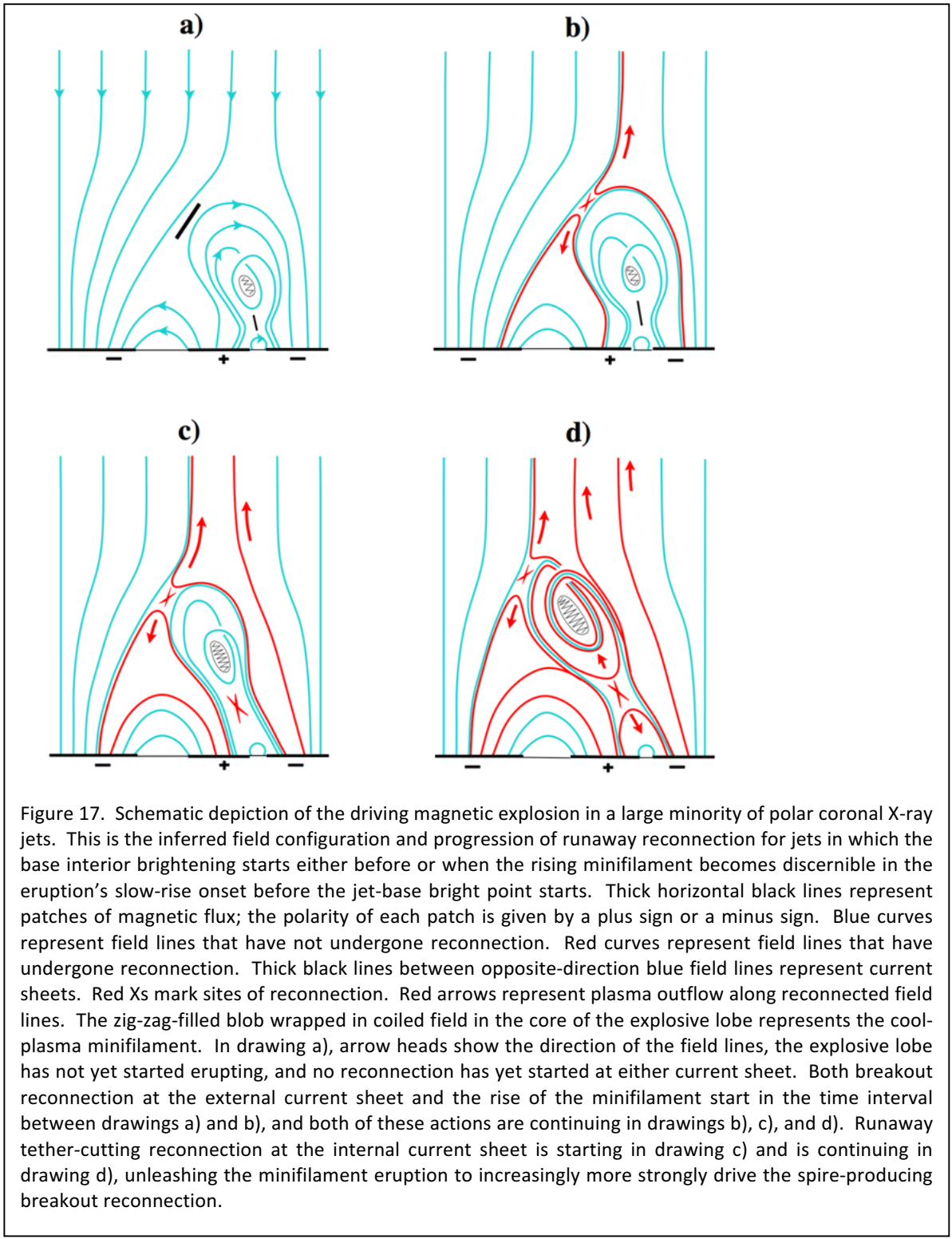

Figure 17. Schematic depiction of the driving magnetic explosion in a large minority of polar coronal X-ray jets. This is the inferred field configuration and progression of runaway reconnection for jets in which the base interior brightening starts either before or when the rising minifilament becomes discernible in the eruption's slow-rise onset before the jet-base bright point starts. Thick horizontal black lines represent patches of magnetic flux; the polarity of each patch is given by a plus sign or a minus sign. Blue curves represent field lines that have not undergone reconnection. Red curves represent field lines that have undergone reconnection. Thick black lines between opposite-direction blue field lines represent current sheets. Red Xs mark sites of reconnection. Red arrows represent plasma outflow along reconnected field lines. The zig-zag-filled blob wrapped in coiled field in the core of the explosive lobe represents the cool-plasma minifilament. In drawing a), arrow heads show the direction of the field lines, the explosive lobe has not yet started erupting, and no reconnection has yet started at either current sheet. Both breakout reconnection at the external current sheet and the rise of the minifilament start in the time interval between drawings a) and b), and both of these actions are continuing in drawings b), c), and d). Runaway tether-cutting reconnection at the internal current sheet is starting in drawing c) and is continuing in drawing d), unleashing the minifilament eruption to increasingly more strongly drive the spire-producing breakout reconnection.

In six jets (Jets 1, 7, 9, 12, 13, 14), the base interior brightening was detected either first (Jets 7, 9, 13) or simultaneously with the first detection of the rising minifilament (Jets 1, 12, 14). In each of the three jets (7, 9, 13) in which the first detection of the rising minifilament was after the first detection of the base interior brightening, nebulous upflow from where the rising minifilament later appeared, upflow along the outside of the stable base arch, is faintly discernible in the AIA movie. This suggests that in these jets, because they were near the limb, the slow rise of the eruption may have started simultaneously with or earlier than the base interior brightening (runaway external reconnection) but the rising minifilament core of the erupting lobe was at first hidden from view by the foreground spicule forest of the chromosphere and cool transition region. (In each of our 15 jets, with or without discernible nebulous upflow before the rising minifilament is first directly detected in the AIA movie, the spicule forest allows the minifilament to have risen a small fraction (~ 0.1) of the height of the base arch before the rising is unambiguously detected in the AIA movie.) In any case, in Table 2, the implication of the base interior brightening starting either first or tied for first with the first detection of the minifilament's slow rise is that at the start of its eruption the explosive lobe filled the void. That is, at the start of the eruption the outside of the explosive lobe was already pressed against the open field on the far side of the stable lobe, making a current sheet there as required for the eruption to have been initiated by runaway external reconnection.

From the above considerations, it is allowed that in a large minority of polar X-ray jets, roughly 40% (6/15), the eruption is initiated by breakout reconnection, which requires the explosive lobe to have been pressed against the opposite open field at eruption onset. The onset and growth of jets that start that way is schematically depicted in Figure 17. The explosive lobe is allowed to have been pressed against the opposite open field at the start of the eruption in any of the other nine of our jets if, for some reason, runaway reconnection at the external current sheet did not start until the eruption was underway and the erupting lobe pressed still harder on the external current sheet. On the other hand, in the 60% (9/15) of our jets in which the rising minifilament is directly detected in the AIA movie before the BIB is detected in the XRT movie, it is allowed that there was initially enough space between the erupting lobe and the open field on the far side of the void in Figure 1 for there to be a null point with no current sheet between the erupting lobe and the open field, and the current sheet was made later by the erupting lobe compressing the field at the null.

## 4. SUMMARY AND DISCUSSION

We have studied the onset of the driving magnetic explosion in 15 random coronal X-ray jets that were observed both by *Hinode*/XRT in coronal X-ray movies of the Sun's polar coronal holes and by *SDO*/AIA in full-disk coronal EUV movies. These jets are 15 of the 20 random polar coronal X-ray jets that Sterling et al (2015) studied and found from an AIA coronal EUV movie of each jet that the jet came from a minifilament eruption. Of these 20 jets, our 15 are those in which the onset of the jet-base-edge bright point (JBP) was observable in the jet's XRT movie.

From each jet's XRT movie and AIA movie, we obtained the time of seven events in the jet's eruption: (1) the first detection of the minifilament rise in the AIA movie, (2) the first detection of the JBP in the XRT movie, (3) the first detection of the base interior brightening (BIB) in the XRT movie, (4) the first detection of the jet's spire in the XRT movie, (5) the first detection of the spire in the AIA movie, (6) the spire maximum in the XRT movie, and (7) the spire maximum in the AIA movie. What we found from the differences between these times can be summarized as follows:

1. In our observations, in each of our jets, the onset of the XRT BIB is either the first sign or the tied-for-first sign of the spire-producing breakout reconnection of the outside of the erupting magnetic arcade that envelops the erupting minifilament flux rope.

2. In a large majority (~ 85%) of polar X-ray jets, the runaway internal tether-cutting reconnection under the erupting minifilament flux rope evidently starts after the start of the breakout reconnection.

3. Among our jets, the start of the minifilament rise relative to the starts of the runaway internal tether-cutting reconnection and the breakout reconnection shows the same diversity as the start of the filament rise in filament eruptions that make a flare and CME and have access to external reconnection. Each of three alternative eruption-onset event sequences is allowed: (1) the minifilament/filament starts rising before there is any evidence of runaway reconnection, either external (breakout) or internal (due to the spicule forest's obscuration of the minifilament at the start of its rise, this option is a possibility for each of our 15 jets); (2) the minifilament/filament starts rising before there is any evidence of runaway internal tether-cutting reconnection but simultaneously with the start of breakout reconnection; (3) the minifilament/filament starts rising simultaneously with the first sign of runaway internal tether-cutting reconnection and before the first sign of breakout reconnection.

4. In correspondence with the fast rise of the filament-carrying erupting arcade turning on with the impulsive burst of runaway internal tether-cutting reconnection in filament eruptions that make a flare and CME, in each of our jets the spire maximum is after the start of the JBP, which is made by runaway internal tether-cutting reconnection under the erupting minifilament flux rope.

5. In a large minority (~ 40% ) of polar X-ray jets, already at the onset of the eruption, there is a current sheet between the explosive arcade and the ambient open field, as we have depicted in Figure 17.

Wyper et al (2017) recently published a numerical MHD simulation of the buildup and eruption of blowout jets that are initiated by breakout reconnection. In that simulation, continual shearing of the field in the explosive lobe builds up a current sheet between the explosive lobe and the open field by inflating the explosive lobe, until the eruption is initiated and unleashed by breakout reconnection at the current sheet, and, as in Figure 17, runaway internal reconnection starts later.

Above results 3 and 4 give further credence to the idea of Sterling et al (2015) that the minifilament-carrying magnetic explosions that make polar coronal X-ray jets work the same way as the filament-carrying larger magnetic explosions that make flares and CMEs. If this idea is valid, then any insight to how the pre-jet minifilament-holding field is built and triggered to erupt should apply to how the pre-eruption filament-holding field is built up and triggered to erupt in a flare and CME.

For each of 10 random coronal jets observed in a quiet region or coronal hole on the central disk (within ~ 45° of disk center), Panesar et al (2016, 2017) study the magnetic-field evolution that gives rise to the minifilament-holding field and triggers that field to erupt. They find that usually the pre-jet minifilament is born no more than a day or two before it erupts in a jet. So, studying the buildup and triggering of the field in pre-jet minifilaments has the advantage that the entire evolution can be followed from before the minifilament's birth through its eruption. In contrast, the entire pre-eruption evolution of a filament field that erupts in a flare and CME usually cannot be followed in comparable entirety and detail because the pre-eruption filament is present for longer than its rotational passage across the central disk, where the evolution of the photospheric magnetic flux can be followed in detail in full-disk magetograms such as those from the Helioseismic and Magnetic Imager (HMI) on *SDO*.

Panesar et al (2016, 2017, 2018) present convincing evidence that the fundamental process that both builds the minifilament field and triggers the magnetic explosion in a coronal jet in a quiet region or coronal hole is usually magnetic flux cancelation at the polarity inversion line under the minifilament. That evidence and the present paper's evidence (that the magnetic explosions in polar coronal X-ray jets are initiated the same way as the magnetic explosions that make a flare and CME, indicating that the magnetic explosion in a polar coronal jet is a small version of the magnetic explosion in a flare and CME) together suggest that usually magnetic flux cancelation inside the arcade that explodes in a flare/CME explosion is the fundamental process that both builds the explosive field and triggers the explosion.

This work was funded by the Heliophysics Division of NASA's Science Mission Directorate through the Living With a Star Targeted Research and Technology Program, the Heliophysics Guest Investigators Program, and the *Hinode* Project. *Hinode* is a Japanese mission developed and launched by ISAS/JAXA, with NAOJ as domestic partner and NASA and STFC (UK) as international partners, and operated by these agencies in co-operation with ESA and NSC (Norway). N.K.P.'s research was supported by an appointment to the NASA Postdoctoral Program at NASA MSFC, administered by the Universities Space Research Association under contract with NASA. R.L.M. benefited from discussions held at the International Space Science Institute (ISSI) in Bern, Switzerland by ISSI's International Team on Pre-eruption Magnetic Configurations of Coronal Mass Ejections (ISSI Team 348). The presentation was improved by helpful comments from the referee.